# Expectation Values of Some Diatomic Molecules With Quantum Interaction Potential In Schrodinger Equation with Hellmann-Feynman Theorem Via Conventional Nikiforov-Uvarov Method.

Ituen B. Okon*[1], Oyebola Popoola[2] and Cecilia N. Isonguyo[3]

1. Theoretical Physics Group, Department of Physics, University of Uyo, Nigeria.
2. Theoretical Physics Group, Department of Physics, University of Ibadan, Nigeria.
3. Theoretical Physics Group, Department of Physics, University of Uyo, Nigeria.

*E-mail of the corresponding author: ituenokon@uniuyo.edu.ng

**Abstract**

In this work, we used a tool of conventional Nikiforov-Uvarov method to determine bound state solution of Schrodinger equation with quantum interaction potential called Hulthen-Yukawa inversely quadratic potential (HYIQP). We obtained the energy eigen values and the total wave function . We employed Hellmann-Feynmann Theorem (HFT) to compute expectation values $<r^{-2}>, <r^{-1}>, <T> \; and <p^{2}>$ for four different diatomic molecules: Hydrogen molecule ($H_2$), Lithium hydride molecule (LiH), Hydrogen Chloride molecule (HCl) and Carbon(II)Oxide molecule. The resulting energy equation reduces to three well known potentials which are: Hulthen potential, Yukawa potential and inversely quadratic potential. We obtained the numerical bound state energies of the expectation values by implementing Matlab algorithm using experimentally determined spectroscopic constant for the different diatomic molecules. We developed a mathematica programming to obtain wave function and probability density plots for different orbital angular quantum number.

## 1.0 **INTRODUCTION**

The study of diatomic molecules is very significant and applicable in many areas of chemical and physical sciences. The excitation of atoms of some diatomic molecules especially the homonuclear diatomic molecules is the principle used in spectrophotometric technique. Diatomic molecules contains two atoms per molecule and can either be homonuclear if it contains two atoms of the same kind or heteronuclear if its contains two atoms of different kind [1-5]. Bound state solutions of relativistic and non-relativistic wave equation arouse a lot of interest for decades. Schrodinger wave equations constitute non-relativistic wave equation while Klein-Gordon Dirac equation constitute the relativistic wave equations. [6-10]. Bound state solutions predominantly have negative energies because the energy of the particle is less than the maximum potential energy [11]. The quantum interaction potential (HYIQP) can be used to compute the bound state energies for both homonuclear and heteronuclear diatomic molecules. Other potentials have been used in studying bound state solutions like: Hulthen, Poschl-Teller, Eckart, Coulomb, Hyllearraas, Pseudoharmonic, scarf II potentials and many others [12-20]. These potentials are studied with some specific methods and techniques such as: Asymptotic iteration method, Nikiforov-Uvarov method, Supersymmetric quantum mechanics approach, formular method, exact quantisation and many more [20-30]. This article is divided into seven sections. Section 1 is the introduction, section 2 is the brief introduction of conventional Nikiforov-Uvarov method. In section 3, we present the radial solution to Schrodinger equation using the proposed potential and obtained both the energy eigen value and their corresponding wave function. In section 4, we have deductions of three well known potential from the proposed potential and compared it to the result of existing literature. In section 5, we introduced HFT and used it to compute the expectation values. In section 6, we present numerical results of the expectation values by implementing Matlab algorithm. Section 7 gives the general conclusion of the article.

The proposed quantum interaction potential is given by

$$V(r) = -\frac{V_0 e^{-2\alpha r}}{\left(1-e^{-2\alpha r}\right)} - \frac{Ae^{-\alpha r}}{r} + \frac{B}{r^2} + C \qquad (1)$$

Where $V_0$ is the potential depth, $\alpha$ is the adjustable parameter known as the screening parameter. A, B and C are all spectroscopic constants which are molecular bond length, molecular constant and potential range respectively..

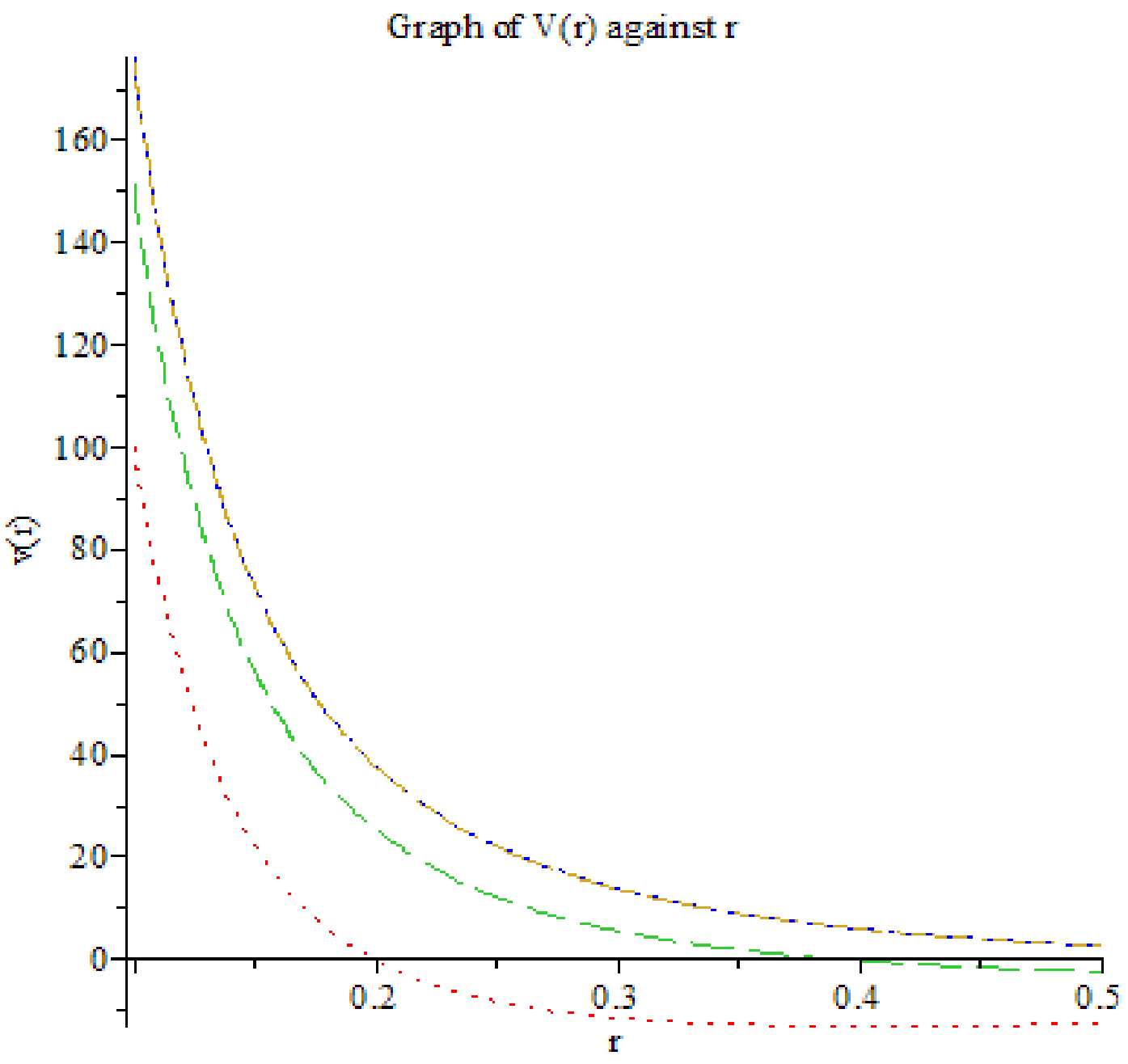

Figure 1: The graph of the quantum interaction potential (HYIQP) for various value of $\alpha$ .

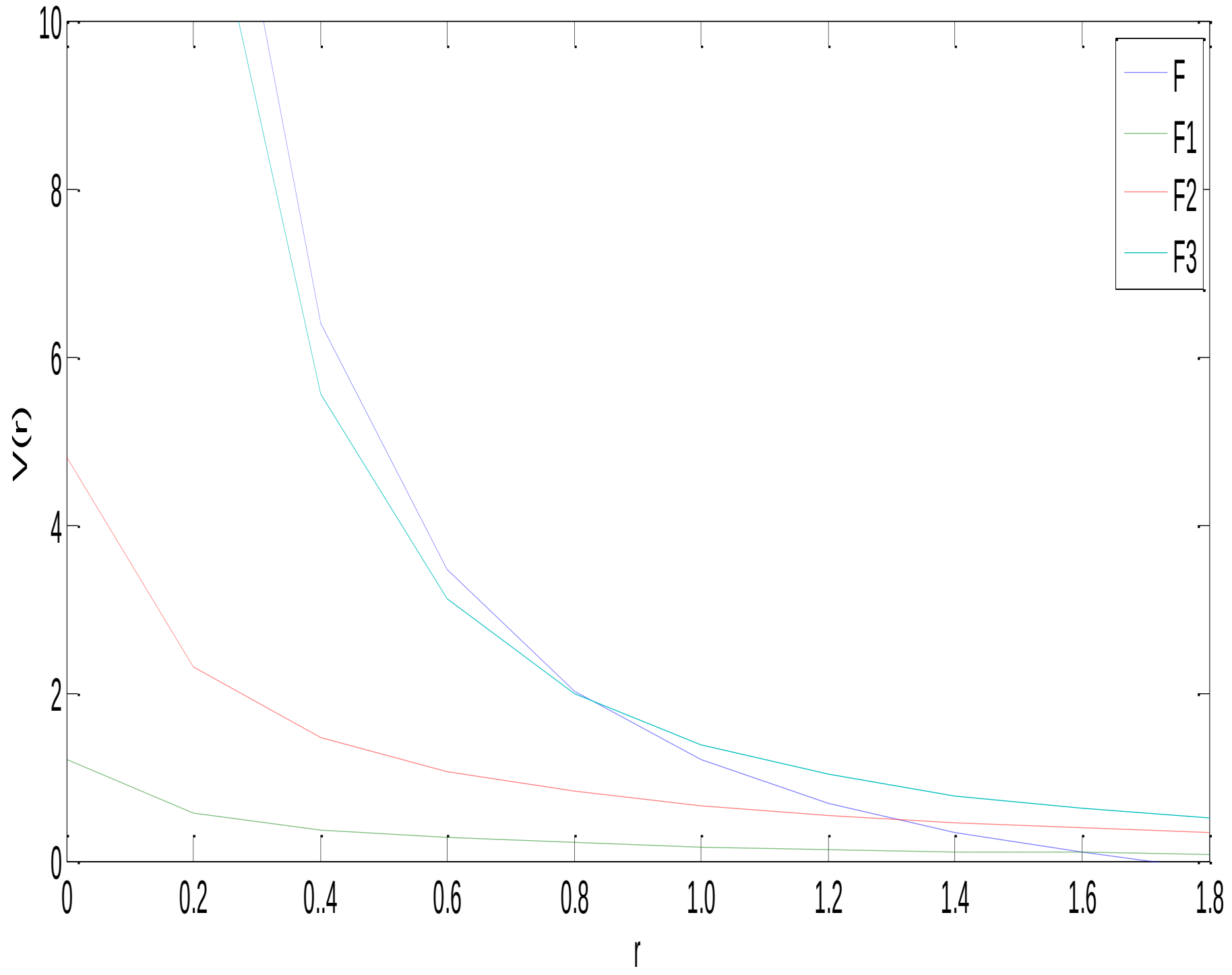

Figure 2. The graph of the HYIQP as compared  to that of individual potentials.

F= HYIQP potential, F1=Hulthen potential, F2= Yukawa potential, F3= inversely quadratic potential. Figure 1 shows the graph of quantum interaction potential for various values of the screening parameter $\alpha$ which decays exponentially. Meanwhile, figure 2 shows the graph of individual potentials plotted in the same scale with the quantum interaction potential(HYIQP). From this Matlab plot, it can be seen that HYIQP is the best suitable in describing bound state energies of diatomic molecules. In elementary quantum mechanics, the wave function, implicitly descibed the behaviour of quantum mechanical systems. We developed this potential model to study the behaviour of four diatomic molecules namely: Hydrogen , Carbon (II) oxide, Lithium hydride and hydrogen Chloride molecules. The wave function and probability density plots, gotten from mathematica programming as seen in figures 3 to 8 below for wave function plots and figure 9 for probability density plots for orbital angular quantum numbers $l = 0,1,2,3,4 and 5$ respectively is very significant in ivestigating the behaviour of diatomic molecules. These plot shows that Carbon (II) Oxide with maximum peak is very flammable and a highly toxic gas though industrially it is a good reducing agent. Lithium hydride possess similar characteristic being a toxic and posionous gas .However, the graph shows that hydrogen chloride and hydrogen molecules are less posionous as compared to Lithium hydride and Carbon(II) oxide molecules.

## SECTION 2: CONVENTIONAL NIKIFOROV-UVAROV METHOD

The NU method is based on reducing second order linear differential equation to a generalized equation of hyper-geometric type[31-32] .This method provides exact solutions in terms of special orthogonal functions as well as corresponding energy eigen value. The NU method is applicable to both relativistic and non-relativistic equations. With appropriate coordinate transformation $S = S(x)$ the equation can be written as

$$\psi''(s)+\frac{\tilde{\tau}(s)}{\sigma(s)}\psi'(s)+\frac{\tilde{\sigma}(s)}{\sigma^2(s)}\psi(s)=0 \qquad (2)$$

$\tilde{\tau}(s)$ is a polynomial of degree one while $\sigma(s)$ and $\tilde{\sigma}(s)$ are polynomials of at most degree two.

In order to find the exact solution to equation (2), we set the wave function as

$$\Psi(s)=\phi(s)\chi(s) \qquad (3)$$

substituting, equation (3) into equation (2) reduces equation (2) into hyper-geometric type.

$$\sigma(s)\chi''(s)+\tau(s)\chi'(s)+\lambda\chi(s)=0 \qquad (4)$$

where the wave function ψ(s) is defined as the logarithmic derivative

$$\frac{\phi^1(s)}{\phi(s)}=\frac{\pi(s)}{\sigma(s)} \qquad (5)$$

Where π(s) is at most polynomial of degree one.
Likewise, the hyper-geometric type function χ(s) in equation (4) for a fixed n is given by the Rodrigues relation

$$\chi_{\text{n}}(s)=\frac{B_n}{\rho(s)}\frac{d^n}{ds^n}\left[\sigma''(s)\rho(s)\right] \qquad (6)$$

$B_n$ is the normalization constant and the weight function ρ(s) satisfy the condition

$$\frac{d}{ds}(\sigma(s)\rho(s))=\tau(s)\rho(s) \qquad (7)$$

Such that

$$\tau(s)=\tilde{\tau}(s)+2\pi(s) \qquad (8)$$

In order to accomplish the conditions imposed on the weight function ρ(s), it is necessary that the classical orthogonal polynomials τ(s) be equal to zero and its derivative be less than zero that is

$$\frac{d\tau(s)}{ds}<0 \qquad (9)$$

Therefore, the function π(s) and the parameter λ required for the NU-method are defined as follows:

$$\pi(s)=\frac{\sigma'-\tilde{\tau}}{2}\pm\sqrt{\left(\frac{\sigma'-\tilde{\tau}}{2}\right)^2-\tilde{\sigma}+k\sigma} \qquad (10)$$

$$\lambda=k+\pi'(s) \qquad (11)$$

The k-values in equation (10) are possible to evaluate if the expression under the square root must be square of polynomials. This is possible, if and only if its discriminant is zero. With this, a new eigen-value equation becomes

$$\lambda = \lambda_n = \frac{nd\tau}{ds} - \frac{n(n-1)}{2}\frac{d^2\sigma}{ds^2}, \text{ n = 0, 1, 2 ...} \qquad (12)$$

where τ(s) is as defined in equation (8) and on comparing equation (11) and equation (12), we obtain the energy Eigen values.

**SECTION 3: RADIAL SOLUTION OF SCHRODINGER EQUATION**

The Schrodinger wave equation as given by

$$\frac{d^2\psi(r)}{dr^2} + \frac{2\mu}{\hbar^2}\left[E - V(r) - \frac{\hbar^2 l(l+1)}{2\mu r^2}\right]\psi(r) = 0 \qquad (13)$$

Substituting equation (1) into (13) gives

$$\frac{d^2\psi(r)}{dr^2} + \frac{2\mu}{\hbar^2}\left[E + \frac{V_0 e^{-2\alpha r}}{\left(1-e^{-2\alpha r}\right)} + \frac{Ae^{-\alpha r}}{r} - \frac{B}{r^2} - C - \frac{\hbar^2 l(l+1)}{2\mu r^2}\right]\psi(r) = 0 \qquad (14)$$

Equation (14) can only be solved analytically to obtain exact solution if the angular orbital momentum number $l = 0$. However, for $l > 0$ equation (14) can only be solve by using some approximations to the centrifugal term. Greene Aldrich approximation is best suitable for equation (14).

Let's define Greene Aldrich approximation as

$$\frac{1}{r^2} = \frac{4\alpha^2 e^{-2\alpha r}}{\left(1-e^{-2\alpha r}\right)^2} \text{ such that } \frac{1}{r} = \frac{2\alpha\ e^{-\alpha r}}{\left(1-e^{-2\alpha r}\right)} \qquad (15)$$

Substituting equation (15) into equation (14) with the transformation $s = e^{-2\alpha r}$ gives

$$\frac{d^2\psi(s)}{ds^2} + \frac{(1-s)}{s(1-s)}\frac{d\psi(s)}{ds} + \frac{1}{s^2(1-s)^2}\begin{bmatrix} -\varepsilon^2(1-s)^2 + \delta^2 s(1-s) + \sigma_1 s(1-s) - \sigma_2 s \\ -\sigma_3(1-s)^2 - sl(l+1) \end{bmatrix}\psi(s) = 0 \qquad (16)$$

Where

$$[\ \varepsilon^2 = -\frac{\mu E}{2\hbar^2\alpha^2},\ \delta^2 = \frac{\mu V_0}{2\hbar^2\alpha^2},\ \sigma_1 = \frac{\mu A}{\hbar^2\alpha},\ \sigma_2 = \frac{2\mu B}{\hbar^2},\ \sigma_3 = \frac{\mu c}{2\hbar^2\alpha^2}\ ] \qquad (17)$$

Simplifying equation (16) further reduced to

$$\frac{d^2\psi(s)}{ds^2} + \frac{(1-s)}{s(1-s)}\frac{d\psi(s)}{ds} + \frac{1}{s^2(1-s)^2}\begin{bmatrix} -(\varepsilon^2+\delta^2+\sigma_1+\sigma_3)s^2 + (2\varepsilon^2+\delta^2+\sigma_1-\sigma_2 \\ +2\sigma_3 - l(l+1))s - (\varepsilon^2+\sigma_3) \\ \end{bmatrix}\psi(s) = 0 \qquad (18)$$

Comparing equation (18) to equation (2) we obtained the followings:

$$
\begin{aligned}
&\sigma(s) = s(1-s) \\
&\tilde{\sigma}(s) = -(\varepsilon^2 + \delta^2 + \sigma_1 + \sigma_3)s^2 + (2\varepsilon^2 + \delta^2 + \sigma_1 - \sigma_2 + 2\sigma_3 - l(l+1))s - (\varepsilon^2 + \sigma_3) \\
&\tilde{\tau}(s) = (1-s)
\end{aligned} \tag{19}
$$

Then, using equation (10), the polynomial equation $\pi(s)$ then become

$$
\pi(s) = \frac{-s}{2} \pm \sqrt{\begin{array}{l} \left(\frac{1}{4} + \varepsilon^2 + \delta^2 + \sigma_1 + \sigma_3 - k\right)s^2 \\ -\left(2\varepsilon^2 + \delta^2 + \sigma_1 - \sigma_2 + 2\sigma_3 - l(l+1) - k\right)s - (\varepsilon^2 + \sigma_3) \end{array}} \tag{20}
$$

To find the value of $k$ we consider the discriminant such that $b^2 - 4ac = 0$

$$
\begin{aligned}
&a = \left(\frac{1}{4} + \varepsilon^2 + \delta^2 + \sigma_1 + \sigma_3 - k\right), \\
&b = -\left(2\varepsilon^2 + \delta^2 + \sigma_1 - \sigma_2 + 2\sigma_3 - l(l+1) - k\right), c = -(\varepsilon^2 + \sigma_3)
\end{aligned} \tag{21}
$$

Hence,

$$
k_1 = -\left(\sigma_2 - \sigma_1 - \delta^2 + l(l+1)\right) + \sqrt{\left(\varepsilon^2 + \sigma_3\right)(4\sigma_2 + 4l(l+1) + 1)} \tag{22}
$$

$$
k_2 = -\left(\sigma_2 - \sigma_1 - \delta^2 + l(l+1)\right) - \sqrt{\left(\varepsilon^2 + \sigma_3\right)(4\sigma_2 + 4l(l+1) + 1)}
$$

Substituting the values of $k_1$ and $k_2$ into equation (20), then the four values of $\pi(s)$ is given below.

$$
\pi(s) = \frac{-s}{2} \pm \left[\begin{array}{ll} \left(\frac{1}{2}\sqrt{(4\sigma_2 + 4l(l+1)+1)} + \sqrt{\left(\varepsilon^2 + \sigma_3\right)}\right)s - \sqrt{\left(\varepsilon^2 + \sigma_3\right)}, & k_1 = -\left(\sigma_2 - \sigma_1 - \delta^2 + l(l+1)\right) + \sqrt{\left(\varepsilon^2 + \sigma_3\right)(4\sigma_2 + 4l(l+1)+1)} \\ \\ \left(\frac{1}{2}\sqrt{(4\sigma_2 + 4l(l+1)+1)} - \sqrt{\left(\varepsilon^2 + \sigma_3\right)}\right)s - \sqrt{\left(\varepsilon^2 + \sigma_3\right)} & k_2 = -\left(\sigma_2 - \sigma_1 - \delta^2 + l(l+1)\right) - \sqrt{\left(\varepsilon^2 + \sigma_3\right)(4\sigma_2 + 4l(l+1)+1)} \end{array}\right] \tag{22a}
$$

$\pi(s)$ has four solutions and one of the solutions satisfied bound state condition which is

$$
\pi(s) = -\frac{s}{2} - \left[\left(\frac{1}{2}\sqrt{(4\sigma_2 + 4l(l+1)+1)} - \sqrt{\left(\varepsilon^2 + \sigma_3\right)}\right)s - \sqrt{\left(\varepsilon^2 + \sigma_3\right)}\right] \tag{22b}
$$

$\frac{d\tau(s)}{ds} < 0$ Is the condition for bound state solution. Using equation (8) we have it that

$$\tau(s) = 1 - 2s - 2\left[\left(\frac{1}{2}\sqrt{(4\sigma_2 + 4l(l+1)+1)} - \sqrt{\left(\varepsilon^2 + \sigma_3\right)}\right)s - \sqrt{\left(\varepsilon^2 + \sigma_3\right)}\right] \tag{23}$$

Such that

$$\tau'(s) = -2 - 2\left[\left(\frac{1}{2}\sqrt{(4\sigma_2 + 4l(l+1)+1)} - \sqrt{\left(\varepsilon^2 + \sigma_3\right)}\right)\right] < 0 \tag{24}$$

Which satisfies the bound state condition. However using equation (11)

$$\lambda = k + \pi'(s) = -\frac{1}{2} - \left(\frac{1}{2}\sqrt{(4\sigma_2 + 4l(l+1)+1)} - \sqrt{\left(\varepsilon^2 + \sigma_3\right)}\right) + \sqrt{\left(\varepsilon^2 + \sigma_3\right)(4\sigma_2 + 4l(l+1)+1)} - \left(\sigma_2 - \sigma_1 - \delta^2 + l(l+1)\right) \tag{25}$$

Using equation (12)

$$\lambda_n = n^2 + n + n\sqrt{(4\sigma_2 + 4l(l+1)+1)} - 2n\sqrt{\left(\varepsilon^2 + \sigma_3\right)} \tag{26}$$

Equating equation (25) and (26) gives the result

$$\varepsilon^2 = \left[\frac{(n^2 + n + \frac{1}{2}) + (n + \frac{1}{2})\sqrt{(4\sigma_2 + 4l(l+1)+1)} + \left(\sigma_2 - \sigma_1 - \delta^2 + l(l+1)\right)}{\left(2n + 1 + \sqrt{(4\sigma_2 + 4l(l+1)+1)}\right)}\right]^2 - \sigma_3 \tag{27}$$

Substituting parameters of equation (17) into equation (27) gives

$$E_{nl} = -\frac{2\hbar^2\alpha^2}{\mu}\left[\frac{\left(\frac{2\mu B}{\hbar^2} - \frac{\mu A}{\hbar^2\alpha} - \frac{\mu V_0}{2\hbar^2\alpha^2} + l(l+1)\right) + (n^2 + n + \frac{1}{2}) + (n + \frac{1}{2})\sqrt{(\frac{8\mu B}{\hbar^2} + 4l(l+1)+1)}}{\left(1 + 2n + \sqrt{(\frac{8\mu B}{\hbar^2} + 4l(l+1)+1)}\right)}\right]^2 + c \tag{28}$$

Equation (28) is the energy for the combined potential.

## 2.1 CALCULATION OF THE WAVE FUNCTION

By using equation (5)

$$\frac{\phi^1(s)}{\phi(s)} = \frac{\pi(s)}{\sigma(s)} \qquad (5)$$

but $\sigma(s) = s - s^2$ and $\pi(s) = -\frac{s}{2} - \left[\left(\frac{1}{2}\sqrt{(4\sigma_2 + 4l(l+1)+1)} + \sqrt{\left(\varepsilon^2 + \sigma_3\right)}\right)s - \sqrt{\left(\varepsilon^2 + \sigma_3\right)}\right]$ then

$$\frac{\pi(s)}{\sigma(s)} = -\frac{1}{2(1-s)} - \frac{1}{s(1-s)}\left[\left(\frac{1}{2}\sqrt{(4\sigma_2 + 4l(l+1)+1)} + \sqrt{\left(\varepsilon^2 + \sigma_3\right)}\right)s - \sqrt{\left(\varepsilon^2 + \sigma_3\right)}\right] \qquad (29)$$

$$\frac{\phi^1(s)}{\phi(s)} = \frac{\pi(s)}{\sigma(s)} = -\frac{1}{2(1-s)} - \frac{1}{s(1-s)}\left[\left(\frac{1}{2}\sqrt{(4\sigma_2 + 4l(l+1)+1)} + \sqrt{\left(\varepsilon^2 + \sigma_3\right)}\right)s - \sqrt{\left(\varepsilon^2 + \sigma_3\right)}\right] \qquad (30)$$

$$\frac{\phi^1(s)}{\phi(s)} = -\frac{1}{2(1-s)} - \frac{1}{s(1-s)}\left[\left(\frac{1}{2}\sqrt{(4\sigma_2 + 4l(l+1)+1)} + \sqrt{\left(\varepsilon^2 + \sigma_3\right)}\right)s - \sqrt{\left(\varepsilon^2 + \sigma_3\right)}\right] \qquad (31)$$

Taking integral of equation (31) gives

$$\int \frac{\phi^1(s)}{\phi(s)}ds = -\frac{1}{2}\int \frac{1}{(1-s)}ds - \frac{1}{2}\sqrt{(4\sigma_2 + 4l(l+1)+1)}\int \frac{1}{(1-s)}ds - \sqrt{\left(\varepsilon^2 + \sigma_3\right)}\int \frac{1}{(1-s)}ds$$
$$+\sqrt{\left(\varepsilon^2 + \sigma_3\right)}\int\left[\frac{1}{s} + \frac{1}{(1-s)}\right]ds \qquad (32)$$

Which then result to

$$\ln\phi(s) = \ln[1-s]^{\frac{1}{2}} + \ln[1-s]^{\frac{1}{2}\sqrt{(4\sigma_2+4l(l+1)+1)}} + \ln[1-s]^{\sqrt{\left(\varepsilon^2+\sigma_3\right)}}$$
$$+\ln[s]^{\sqrt{\left(\varepsilon^2+\sigma_3\right)}} - \ln[1-s]^{\sqrt{\left(\varepsilon^2+\sigma_3\right)}} \qquad (33)$$

Equation (33) can further be reduced to

$$\ln\phi(s) = \ln\left\{[1-s]^{\left(\frac{1}{2}+\frac{1}{2}\sqrt{(4\sigma_2+4l(l+1)+1)}\right)}.[s]^{\sqrt{\left(\varepsilon^2+\sigma_3\right)}}\right\} \qquad (34)$$

The integration constant is ignored since the final equation is to be express in terms of Rodrique relation with normalization constant. Taking exponent of equation (34) gives

$$\phi(s) = [1-s]^{\left(\frac{1}{2}+\frac{1}{2}\sqrt{(4\sigma_2+4l(l+1)+1)}\right)} s^{\sqrt{(\varepsilon^2+\sigma_3)}} \tag{35}$$

Equation (35) gives the first part of the wave function. To determine the second part of the wave function, we first of all calculate the weight function.

## 2.2 CALCULATION OF WEIGHT FUNCTION.

Using equation (7)

$$\frac{d}{ds}\left(\sigma(s)\rho(s)\right) = \tau(s)\rho(s) \Rightarrow \int \frac{\rho'(s)}{\rho(s)}ds = \int \frac{\tau(s)-\sigma'(s)}{\sigma(s)}ds$$

Substituting the parameters gives

$$\ln\rho(s) = \int \frac{1-2s-2\left[\left(\frac{1}{2}\sqrt{(4\sigma_2+4l(l+1)+1)}+\sqrt{(\varepsilon^2+\sigma_3)}\right)s\right]+2\sqrt{(\varepsilon^2+\sigma_3)}-1+2s}{s(1-s)}ds \tag{36}$$

Integrating equation (36) gives

$$\rho(s) = [1-s]^{\left(-2\sqrt{(\varepsilon^2+\sigma_3)}-\sqrt{(4\sigma_2+4l(l+1)+1)}\right)} s^{4\sqrt{(\varepsilon^2+\sigma_3)}} \tag{37}$$

Rewriting equation (37) in its Rodrigue form by making use of equation (6) gives

$$\begin{aligned} Y_n(s) &= B_n(s) s^{-4\sqrt{(\varepsilon^2+\sigma_3)}} [1-s]^{-\left(-2\sqrt{(\varepsilon^2+\sigma_3)}-\sqrt{(4\sigma_2+4l(l+1)+1)}\right)} \\ &\times \frac{d^n}{ds^n}\left[ s^{n+4\sqrt{(\varepsilon^2+\sigma_3)}} [1-s]^{n+\left(-2\sqrt{(\varepsilon^2+\sigma_3)}-\sqrt{(4\sigma_2+4l(l+1)+1)}\right)} \right] \end{aligned} \tag{38}$$

Lets define standard associated Laguerre polynomial as

$$\chi_n(s) = B_n(s) s^{-\upsilon}(1-s)^{-\mu} \frac{d^n}{ds^n}\left[ s^{n+\upsilon}(1-s)^{n+\mu} \right] = P_n^{[\mu+\upsilon,\mu]}(1-s) \tag{39}$$

Then re-writing equation (38) in terms of equation (39) gives the second part of the wave function as

$$Y_n(s) = B_n(s) P_n^{\left[\left(2\sqrt{(\varepsilon^2+\sigma_3)}-4\sqrt{(4\sigma_2+4l(l+1)+1)}\right),\left(-2\sqrt{(\varepsilon^2+\sigma_3)}-4\sqrt{(4\sigma_2+4l(l+1)+1)}\right)\right]}(1-s) \tag{40}$$

Hence the total wave function is given by

$$\Psi_n(s)=\phi(s)Y_n(s)=B_n(s)P_n^{\left[\left(2\sqrt{(\varepsilon^2+\sigma_3)}-4\sqrt{(4\sigma_2+4l(l+1)+1)}\right),\left(-2\sqrt{(\varepsilon^2+\sigma_3)}-4\sqrt{(4\sigma_2+4l(l+1)+1)}\right)\right]}(1-s)$$
$$\times[1-s]^{\left(\frac{1}{2}+\frac{1}{2}\sqrt{(4\sigma_2+4l(l+1)+1)}\right)}s^{\sqrt{(\varepsilon^2+\sigma_3)}} \quad (41)$$

Substituting equation (17) into (41) reduced it to

$$\Psi_n(s)=\phi(s)Y_n(s)=B_n(s)P_n^{\left[\left(2\sqrt{\left(\frac{\mu c}{2\hbar^2\alpha^2}-\frac{\mu E}{2\hbar^2\alpha^2}\right)}-4\sqrt{(\frac{8\mu B}{\hbar^2}+4l(l+1)+1)}\right),\left(-2\sqrt{\left(\frac{\mu c}{2\hbar^2\alpha^2}-\frac{\mu E}{2\hbar^2\alpha^2}\right)}-4\sqrt{(\frac{8\mu B}{\hbar^2}+4l(l+1)+1)}\right)\right]}(1-e^{-2\alpha r})$$
$$\times\left[1-e^{-2\alpha r}\right]^{\left(\frac{1}{2}+\frac{1}{2}\sqrt{(\frac{8\mu B}{\hbar^2}+4l(l+1)+1)}\right)}(e^{-2\alpha r})^{\sqrt{\left(\left(\frac{\mu c}{2\hbar^2\alpha^2}-\frac{\mu E}{2\hbar^2\alpha^2}\right)\right)}} \quad (42)$$

Equation (42) is the total wave function for proposed potential. Equation (42) can further be reduced to

$$\Psi_n(s)=\phi(s)Y_n(s)=B_n(s)P_n^{\left[\left(2\sqrt{\left(\frac{\mu c}{2\hbar^2\alpha^2}-\frac{\mu E}{2\hbar^2\alpha^2}\right)}-4\sqrt{(\frac{8\mu B}{\hbar^2}+4l(l+1)+1)}\right),\left(-2\sqrt{\left(\frac{\mu c}{2\hbar^2\alpha^2}-\frac{\mu E}{2\hbar^2\alpha^2}\right)}-4\sqrt{(\frac{8\mu B}{\hbar^2}+4l(l+1)+1)}\right)\right]}(1-s)$$
$$\times[1-s]^{\left(\frac{1}{2}+\frac{1}{2}\sqrt{(\frac{8\mu B}{\hbar^2}+4l(l+1)+1)}\right)}(s)^{\sqrt{\left(\left(\frac{\mu c}{2\hbar^2\alpha^2}-\frac{\mu E}{2\hbar^2\alpha^2}\right)\right)}} \quad (42)$$

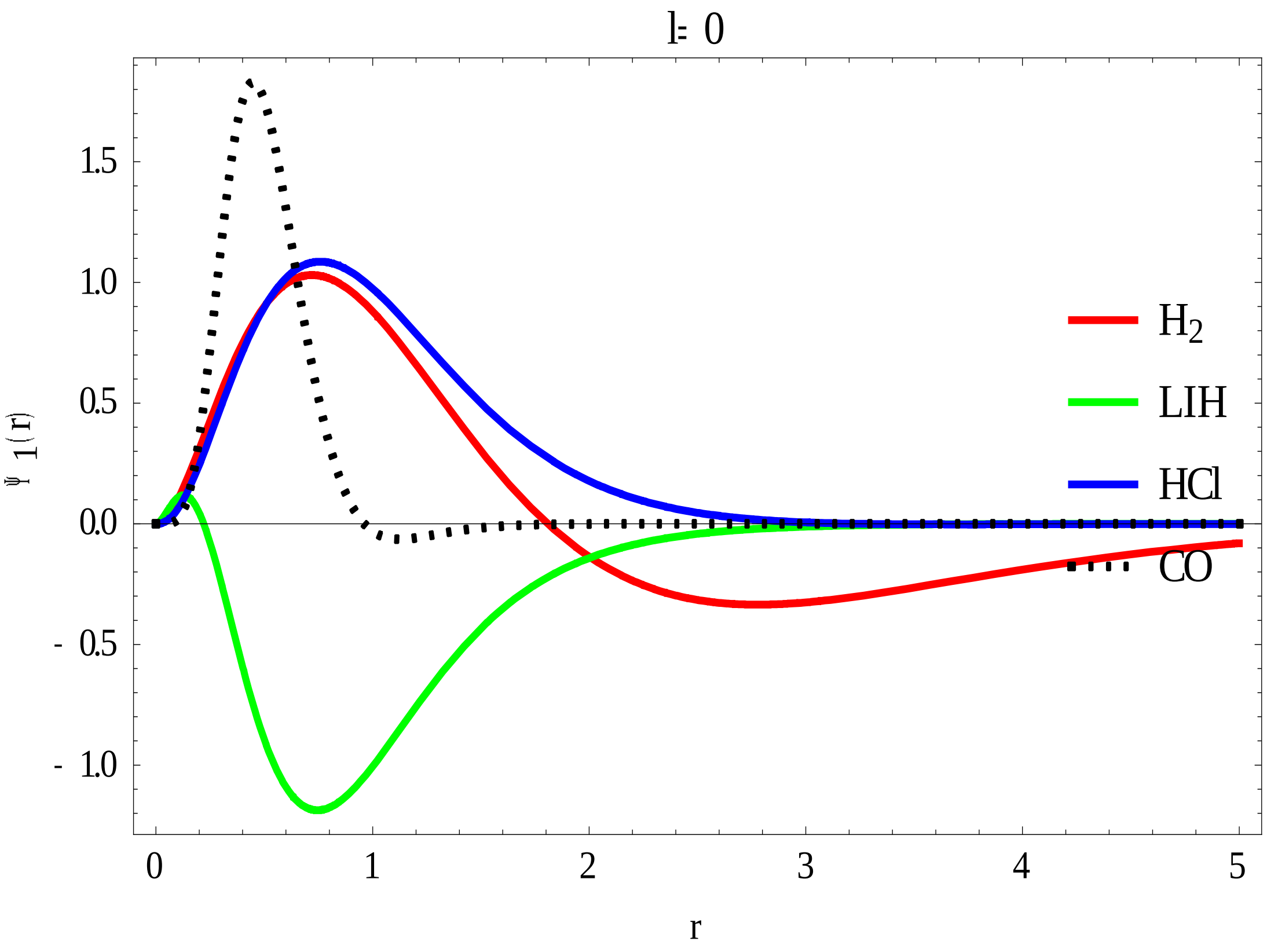

**Fig 3: Wave Function Plot for the Combined Potential with Orbital Angular Quantum Number** $l = 0$

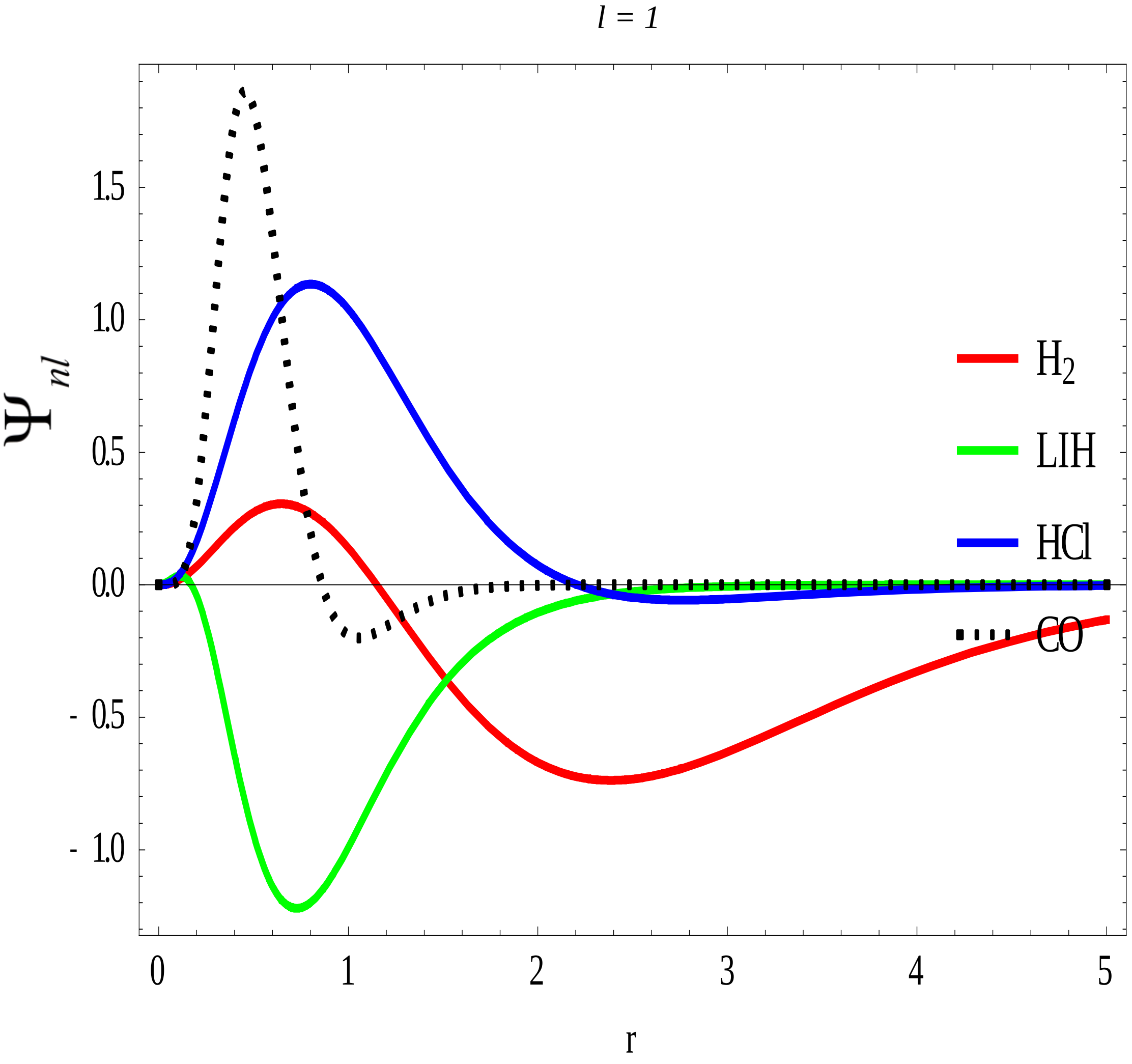

**Fig 4: Wave Function Plot for the Combine Potential with Orbital Angular Quantum Number $l = 1$**

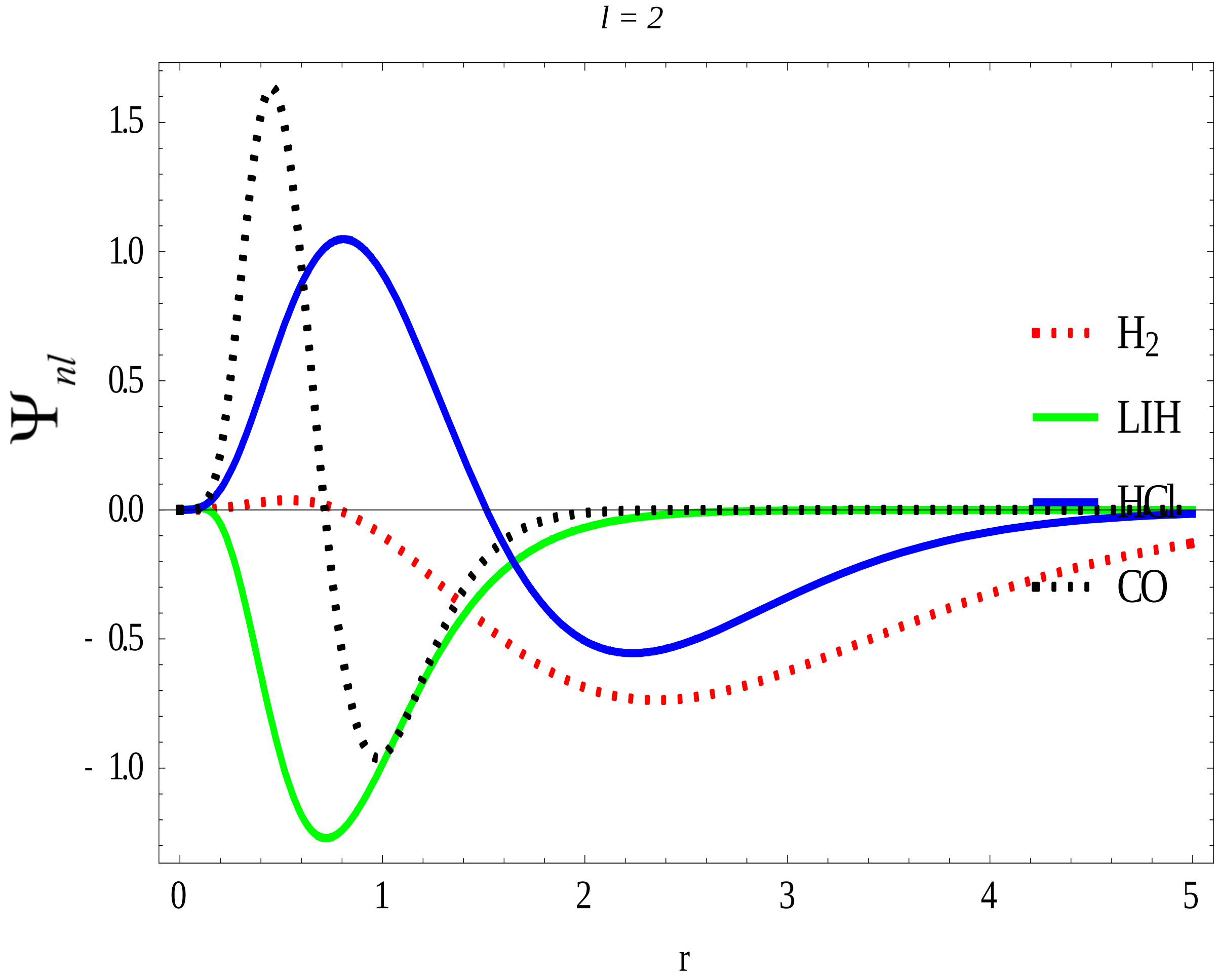

**Fig5: Wave Function Plot for the Combine Potential with Orbital Angular Quantum Number** $l = 2$

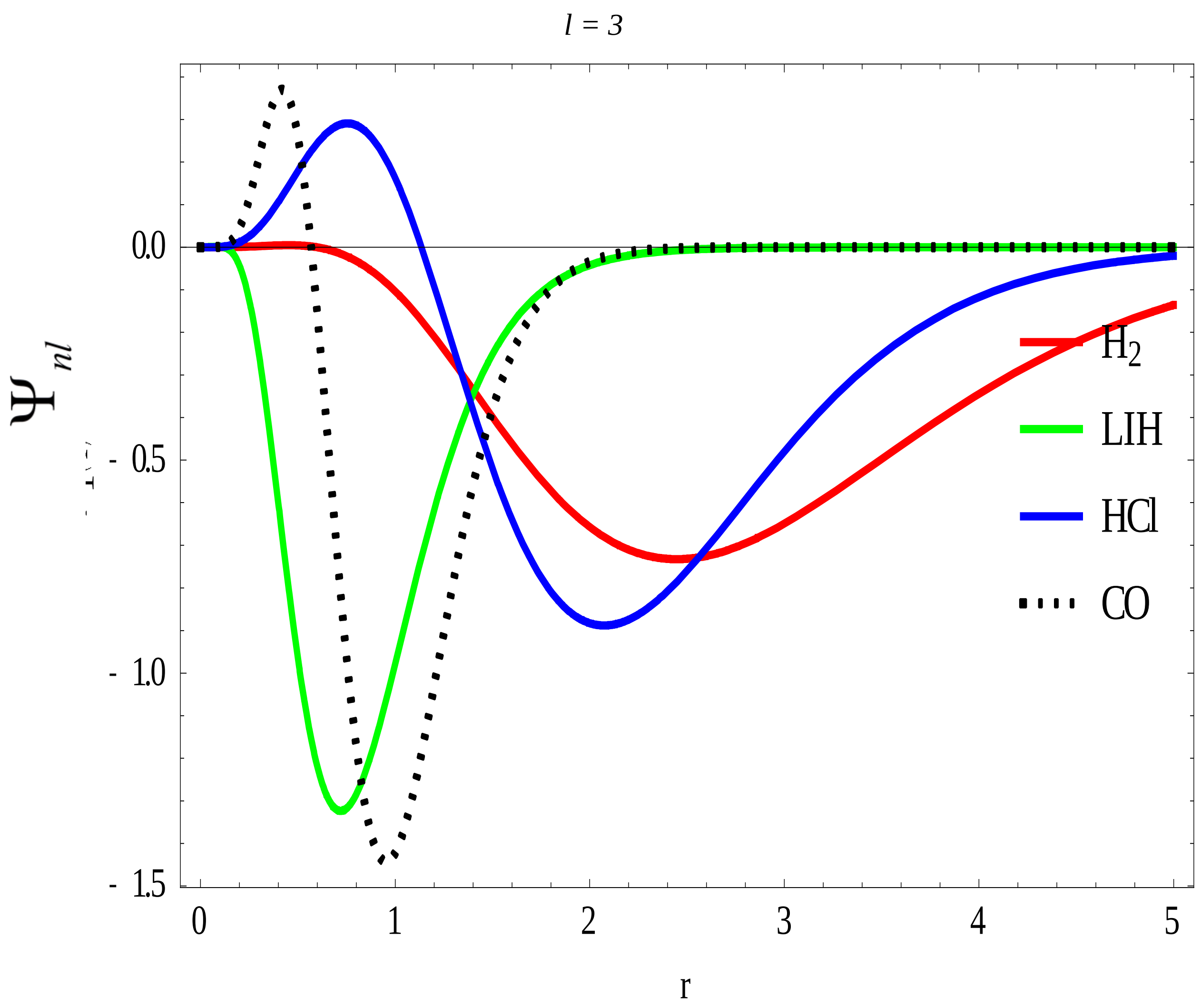

**Fig 6: Wave Function Plot for the Combined Potential with Orbital Angular Quantum Number** $l = 3$

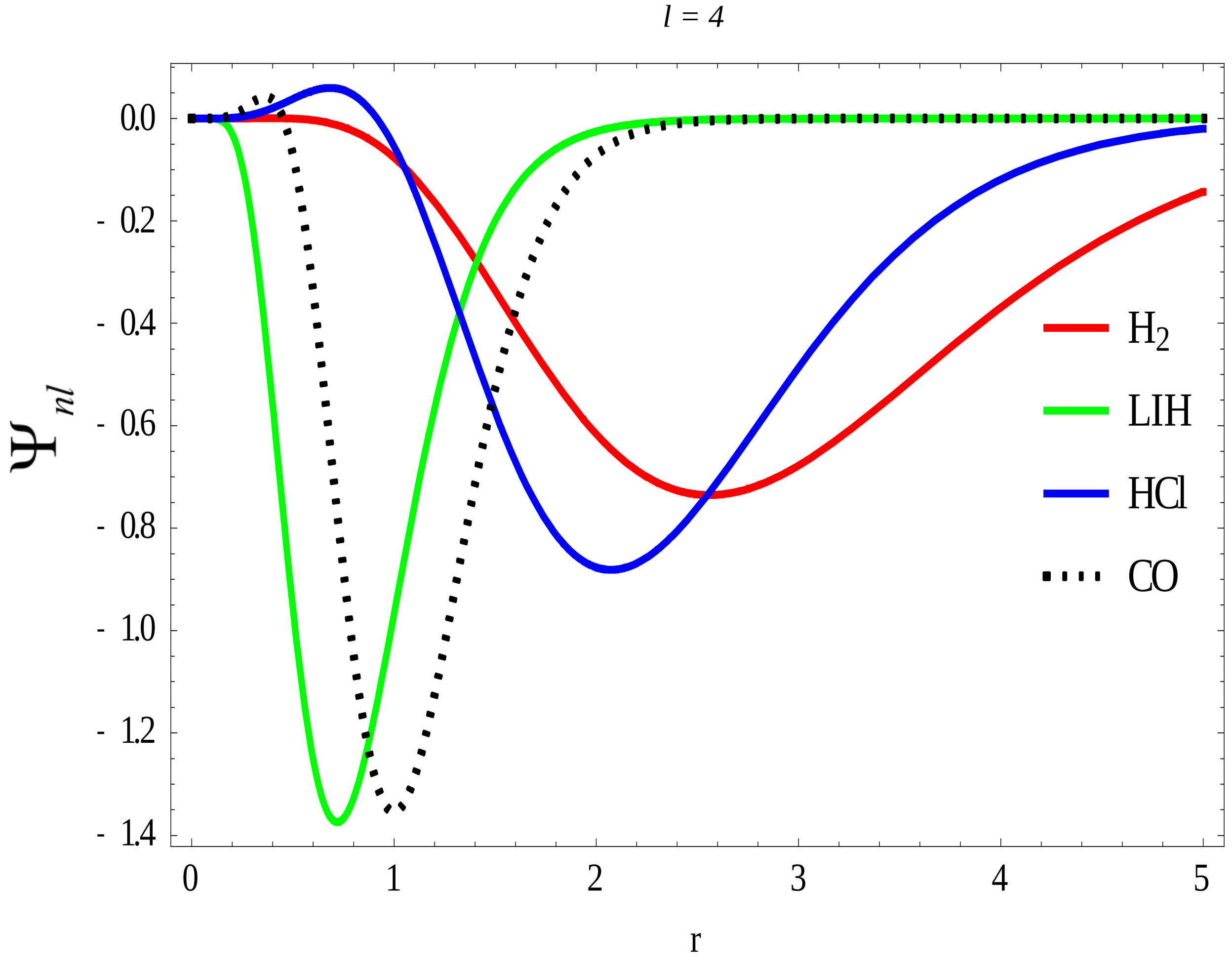

**Fig 7: Wave Function Plot for the Combine Potential with Orbital Angular Quantum Number** $l = 4$

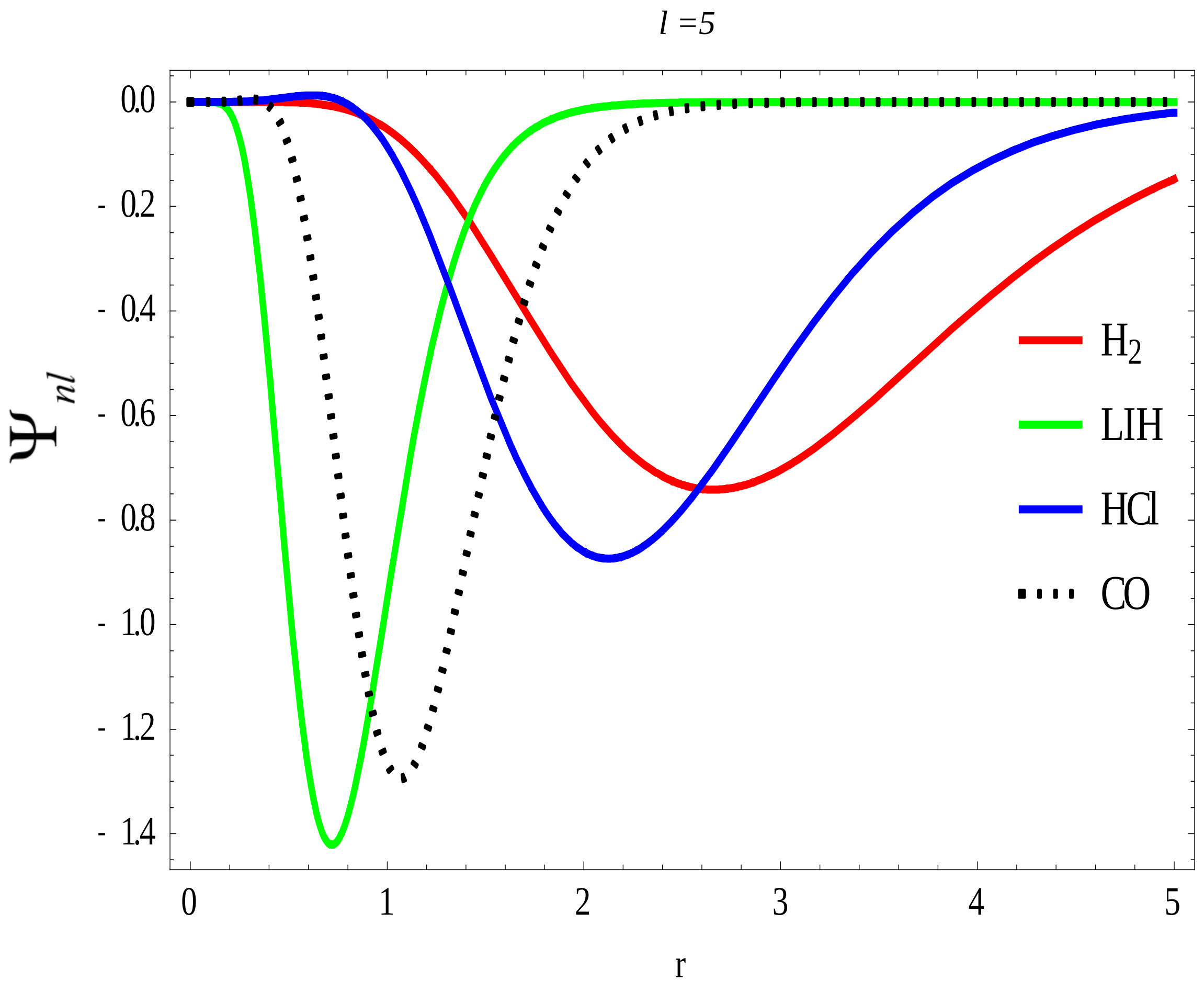

**Fig 8: Wave Function Plot for the Combined Potential with Orbital Angular Quantum Number** $l = 5$

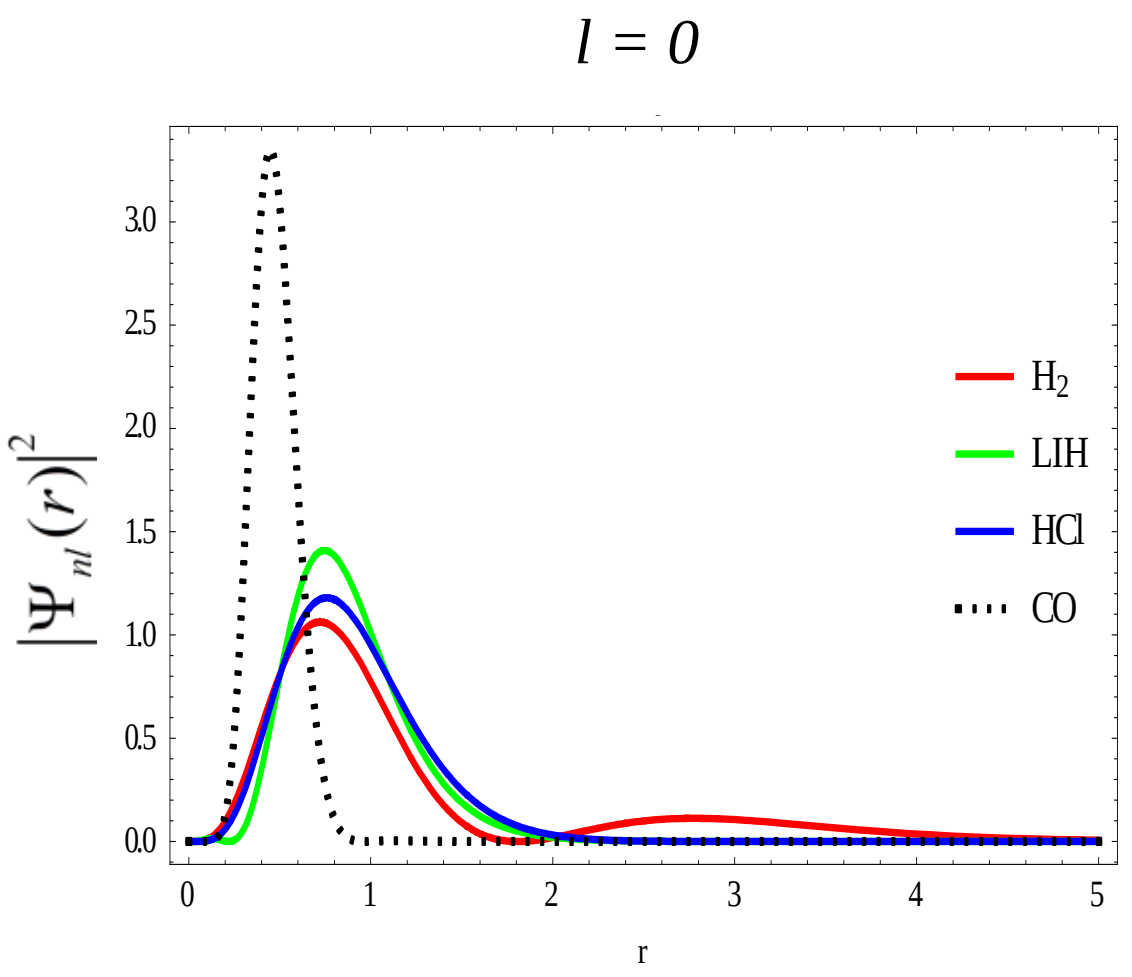

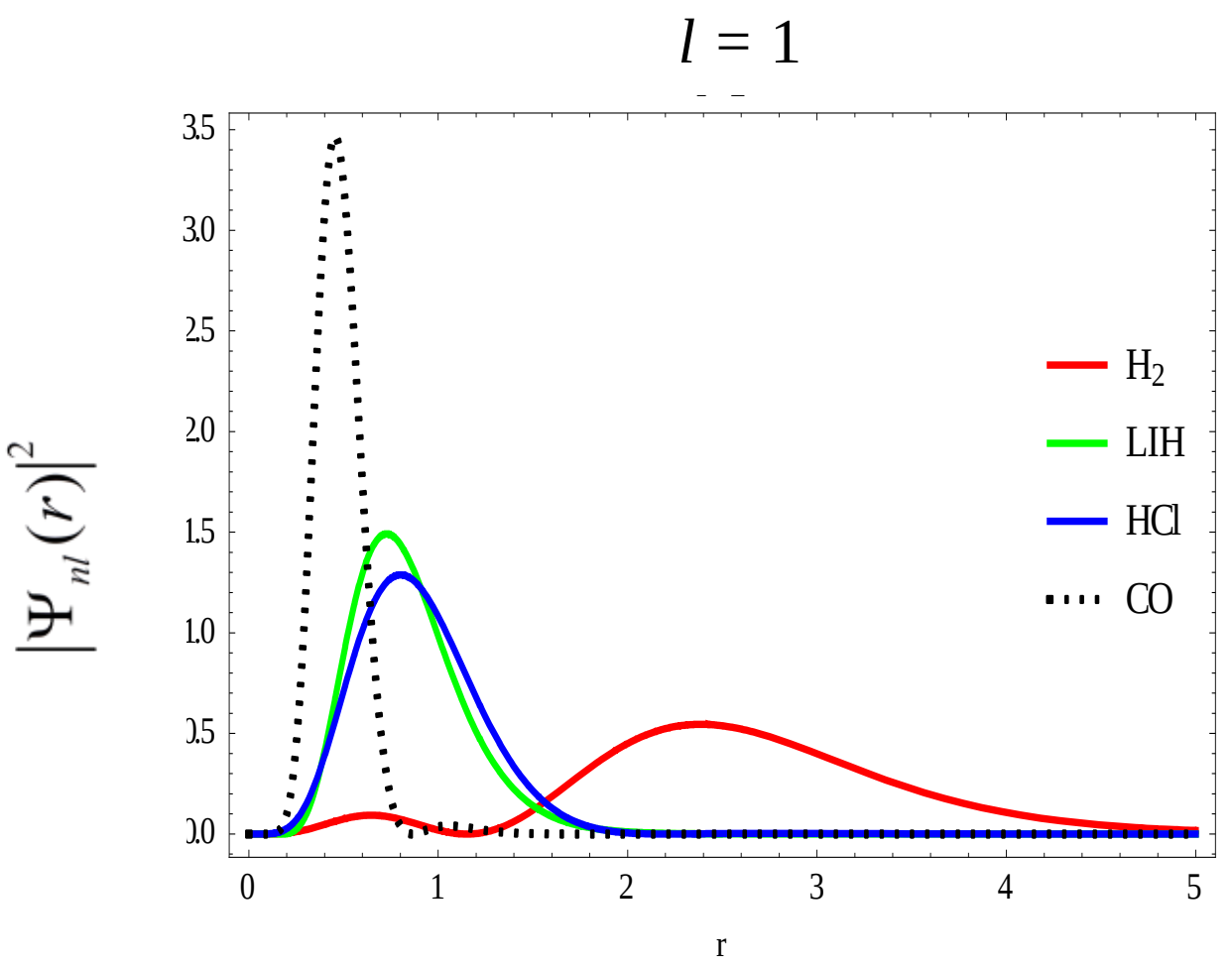

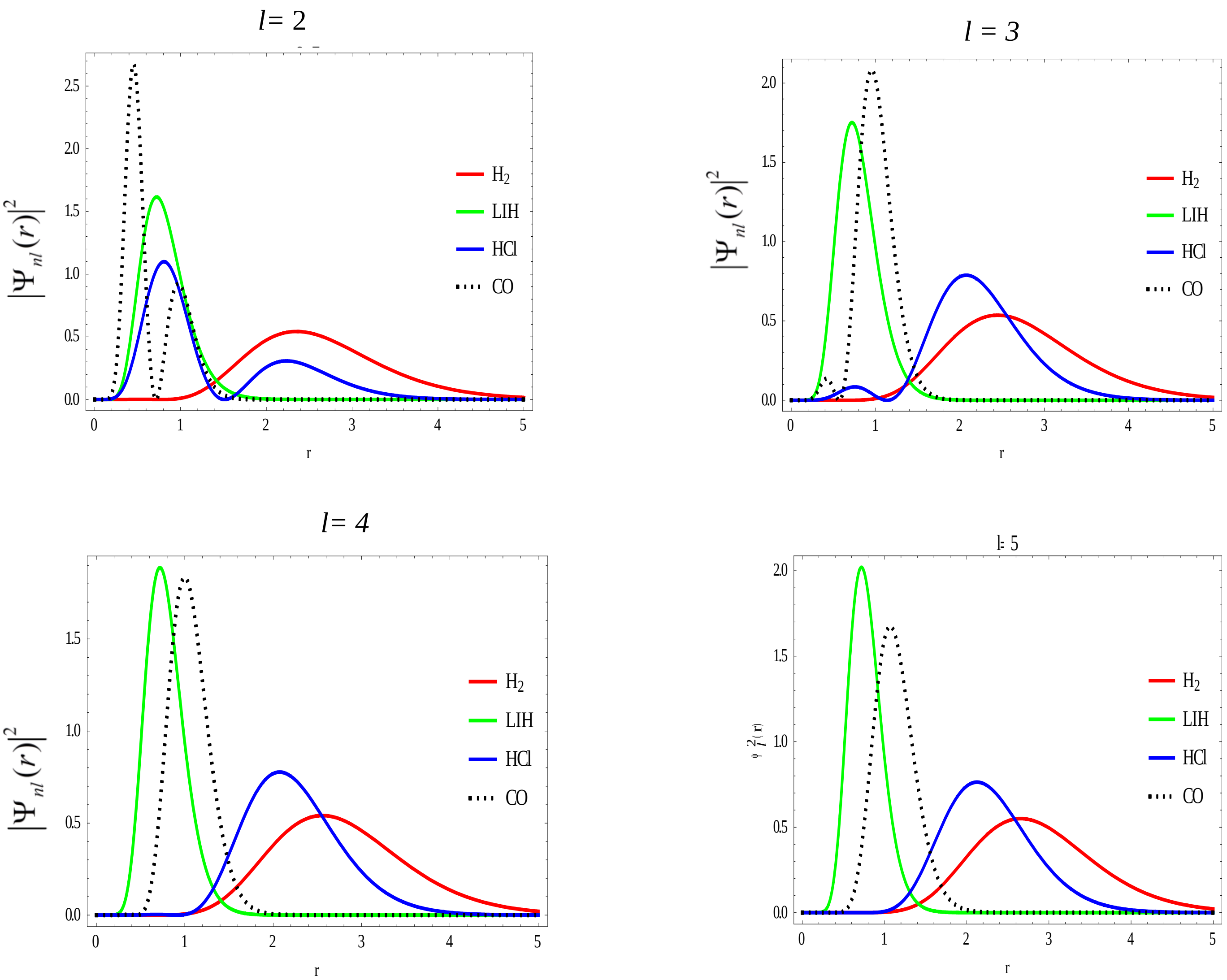

**Fig 9: Probability density Plots for the Combined Potential with Orbital Angular Quantum Number** $l = 0, 1, 2, 3, 4 \, and \, 5$

**SECTION 4: DEDUCTIONS FROM THE PROPOSED POTENTIAL**.

## 4.1 HULTHEN POTENTIAL

If we set A=B=C=0 in equation (1) we obtained Hulthen potential as

$$V(r)=\frac{-V_0e^{-2\alpha r}}{\left(1-e^{-2\alpha r}\right)} \tag{43}$$

The energy of this potential is given as

$$E_{nl}=-\frac{2\hbar^2\alpha^2}{\mu}\left[\frac{\left(-\frac{\mu V_0}{2\hbar^2\alpha^2}+l(l+1)\right)+(n^2+n+\frac{1}{2})+(n+\frac{1}{2})\sqrt{(4l(l+1)+1)}}{\left(1+2n+\sqrt{(4l(l+1)+1)}\right)}\right]^2 \tag{44}$$

However $\sqrt{(4l(l+1)+1)}=2l+1$ , then equation (44) becomes

$$E_{nl}=-\frac{2\hbar^2\alpha^2}{\mu}\left[\frac{\left(-\frac{\mu V_0}{2\hbar^2\alpha^2}\right)+(n+l)(n+l+2)+1}{2\left(n+l+1\right)}\right]^2 \tag{45}$$

**4.2 YUKAWA POTENTIAL**

If we set $V_0=B=C=0$ in equation (1) then the potential reduced to Yukawa potential.

$$V(r)=-\frac{Ae^{-\alpha r}}{r} \tag{46}$$

The corresponding energy equation for this potential is given as

$$E_{nl}=-\frac{2\hbar^2\alpha^2}{\mu}\left[\frac{\left(-\frac{\mu A}{\hbar^2\alpha}+l(l+1)\right)+(n^2+n+\frac{1}{2})+(n+\frac{1}{2})\sqrt{(4l(l+1)+1)}}{\left(1+2n+\sqrt{(4l(l+1)+1)}\right)}\right]^2 \tag{47}$$

**4.3 INVERSELY QUADRATIC POTENTIAL**

If we set $V_0 = A = C = 0$ in equation (1) then the potential reduces to inversely quadratic potential.

$$V(r) = \frac{B}{r^2} \tag{48}$$

The corresponding energy for this potential is given a

$$E_{nl} = -\frac{2\hbar^2\alpha^2}{\mu}\left[\frac{\left(\frac{2\mu B}{\hbar^2} + l(l+1)\right) + (n^2 + n + \frac{1}{2}) + (n + \frac{1}{2})\sqrt{(\frac{8\mu B}{\hbar^2} + 4l(l+1) + 1)}}{\left(1 + 2n + \sqrt{(\frac{8\mu B}{\hbar^2} + 4l(l+1) + 1)}\right)}\right]^2 \tag{49}$$

**SECTION 5: HELLMANN-FEYNMAN THEOREM**

Hellmann-Feynman theorem (HFT) is commonly used in the calculation of intermolecular forces in molecules[33]. However, inorder to engage HFT in calculating the expectation values, one then promote the fixed parameters which appears in the Hamiltonian to be continuous variable in order to ease the mathematical purpose of taking the derivative. This theorem state that if Hamiltonian H for a particular quantum mechanical system is given as a function of some parameters q. Let E (q) and $\Psi(q)$ be the eigen values and the eigen functions of Hamiltonian $H(q)$ respectively. Then:

$$\frac{\partial E_n(q)}{\partial q} = \left\langle \Psi(q) \right| \frac{\partial H(q)}{\partial q} \left| \Psi(q) \right\rangle . \qquad (50)$$

However, the Hamiltonian which contains the effective potential can be express as

$$H = \frac{\hbar^2}{2\mu}\frac{d^2}{dr^2} + \frac{\hbar^2 l(l+1)}{2\mu r^2} - \frac{v_0 e^{-2\alpha r}}{\left(1-e^{-2\alpha r}\right)} - \frac{Ae^{-\alpha r}}{r} + \frac{B}{r^2} + C \qquad (51)$$

Let's recall the energy equation as given in equation (28)

$$E_{nl} = -\frac{2\hbar^2\alpha^2}{\mu}\left[\frac{\left(\frac{2\mu B}{\hbar^2} - \frac{\mu A}{\hbar^2\alpha} - \frac{\mu V_0}{2\hbar^2\alpha^2} + l(l+1)\right) + (n^2+n+\frac{1}{2}) + (n+\frac{1}{2})\sqrt{(\frac{8\mu B}{\hbar^2} + 4l(l+1)+1)}}{\left(1+2n+\sqrt{(\frac{8\mu B}{\hbar^2} + 4l(l+1)+1)}\right)}\right]^2 + c \, (28)$$

**5.1 EXPECTATION VALUE OF $< r^{-2} >$.**

Substituting : $q = l$ into equation (50) then,

$$\frac{\partial E_n(l)}{\partial l} = \left\langle \Psi(l) \right| \frac{\partial H(l)}{\partial l} \left| \Psi(l) \right\rangle \qquad (52)$$

Taking the partial derivative of equation (28) with respect to $l$ gives

$$\frac{\partial E_n(l)}{\partial l}=\frac{-4\hbar^2\alpha^2}{\mu}\left\{\frac{\left[1+2n+\sqrt{\frac{8\mu B}{\hbar^2}+4l(l+1)+1}\right]\left[(2l+1)+(4n+2)(2l+1)\right]-\left[\frac{\left(\frac{2\mu B}{\hbar^2}-\frac{\mu A}{\hbar^2\alpha}-\frac{\mu V_0}{2\hbar^2\alpha^2}+l(l+1)\right)(4l+2)}{\sqrt{\frac{8\mu B}{\hbar^2}+4l(l+1)+1}}+\frac{\left(n^2+n+\frac{1}{2}\right)(4l+2)}{\sqrt{\frac{8\mu B}{\hbar^2}+4l(l+1)+1}}+\left(n+\frac{1}{2}\right)(4l+2)\right]}{\left[1+2n+\sqrt{\frac{8\mu B}{\hbar^2}+4l(l+1)+1}\right]^2}\right\}\times$$

$$\left\{\frac{\left(\frac{2\mu B}{\hbar^2}-\frac{\mu A}{\hbar^2\alpha}-\frac{\mu V_0}{2\hbar^2\alpha^2}+l(l+1)\right)+(n^2+n+\frac{1}{2})+(n+\frac{1}{2})\sqrt{\frac{8\mu B}{\hbar^2}+4l(l+1)+1}}{\left[1+2n+\sqrt{\frac{8\mu B}{\hbar^2}+4l(l+1)+1}\right]}\right\} \tag{53}$$

Taking the partial derivative of equation (51) with respect to $l$ gives

$$\left\langle\Psi(l)\right|\frac{\partial H(l)}{\partial l}\left|\Psi(l)\right\rangle=\frac{\hbar^2}{2\mu}\left(2l+1\right)<r^{-2}>. \tag{54}$$

Equating equation (53) to (54) gives the expectation values of $<r^{-2}>$ for different orbital quantum number. Hence,

$$<r^{-2}>=\frac{-8\alpha^2}{(2l+1)}\left\{\frac{\left[1+2n+\sqrt{\frac{8\mu B}{\hbar^2}+4l(l+1)+1}\right]\left[(2l+1)+(4n+2)(2l+1)\right]-\left[\frac{\left(\frac{2\mu B}{\hbar^2}-\frac{\mu A}{\hbar^2\alpha}-\frac{\mu V_0}{2\hbar^2\alpha^2}+l(l+1)\right)(4l+2)}{\sqrt{\frac{8\mu B}{\hbar^2}+4l(l+1)+1}}+\frac{\left(n^2+n+\frac{1}{2}\right)(4l+2)}{\sqrt{\frac{8\mu B}{\hbar^2}+4l(l+1)+1}}+\left(n+\frac{1}{2}\right)(4l+2)\right]}{\left[1+2n+\sqrt{\frac{8\mu B}{\hbar^2}+4l(l+1)+1}\right]^2}\right\}\times$$

$$\left\{\frac{\left(\frac{2\mu B}{\hbar^2}-\frac{\mu A}{\hbar^2\alpha}-\frac{\mu V_0}{2\hbar^2\alpha^2}+l(l+1)\right)+(n^2+n+\frac{1}{2})+(n+\frac{1}{2})\sqrt{\frac{8\mu B}{\hbar^2}+4l(l+1)+1}}{\left[1+2n+\sqrt{\frac{8\mu B}{\hbar^2}+4l(l+1)+1}\right]}\right\} \qquad (55)$$

**5.2 EXPECTATION VALUE OF $<r^{-1}>$.**

Taking the partial derivative of equation (28) with respect to $A$ gives

$$\frac{\partial E_{nl}}{\partial A}=\frac{4\alpha}{\left[1+2n+\sqrt{\frac{8\mu B}{\hbar^2}+4l(l+1)+1}\right]^2}\left\{\begin{array}{l}\left(\frac{2\mu B}{\hbar^2}-\frac{\mu A}{\hbar^2\alpha}-\frac{\mu V_0}{2\hbar^2\alpha^2}+l(l+1)\right)\\+(n^2+n+\frac{1}{2})+(n+\frac{1}{2})\sqrt{\frac{8\mu B}{\hbar^2}+4l(l+1)+1}\end{array}\right\} \qquad (56)$$

Also, taking the partial derivative of equation (51) with respect to $A$ gives

$$\langle\Psi(A)|\frac{\partial H(A)}{\partial A}|\Psi(A)\rangle=-e^{-\alpha r}<r^{-1}> \qquad (57)$$

Equating equation (56) to (57) then gives

$$<r^{-1}>=\frac{-4\alpha e^{\alpha r}}{\left[1+2n+\sqrt{\frac{8\mu B}{\hbar^2}+4l(l+1)+1}\right]^2}\left\{\begin{array}{l}\left(\frac{2\mu B}{\hbar^2}-\frac{\mu A}{\hbar^2\alpha}-\frac{\mu V_0}{2\hbar^2\alpha^2}+l(l+1)\right)\\+(n^2+n+\frac{1}{2})+(n+\frac{1}{2})\sqrt{\frac{8\mu B}{\hbar^2}+4l(l+1)+1}\end{array}\right\} \qquad (58)$$

**5.3 EXPECTATION VALUES FOR $<T>$ AND $<p^2>$**

Taking the partial derivative of equation (28) with respect to $\mu$ . This then gives

$$
\frac{\partial E_n(\mu)}{\partial \mu}=\frac{-4\hbar^2\alpha^2}{\mu}\left\{\frac{\left[1+2n+\sqrt{\frac{8\mu B}{\hbar^2}+4l(l+1)+1}\right]\left[\frac{2B}{\hbar^2}-\frac{A}{\hbar^2\alpha}-\frac{v_0}{2\hbar^2\alpha^2}+\frac{\left(n+\frac{1}{2}\right)4B}{\hbar^2\sqrt{\frac{8\mu B}{\hbar^2}+4l(l+1)+1}}\right]-\left[\frac{\left(\frac{2\mu B}{\hbar^2}-\frac{\mu A}{\hbar^2\alpha}-\frac{\mu V_0}{2\hbar^2\alpha^2}+l(l+1)\right)4B}{\hbar^2\sqrt{\frac{8\mu B}{\hbar^2}+4l(l+1)+1}}+\frac{\left(n^2+n+\frac{1}{2}\right)(4B)}{\hbar^2\sqrt{\frac{8\mu B}{\hbar^2}+4l(l+1)+1}}+\frac{\left(n+\frac{1}{2}\right)(4B)}{\hbar^2}\right]}{\left[1+2n+\sqrt{\frac{8\mu B}{\hbar^2}+4l(l+1)+1}\right]^2}\right\}\times
$$

$$
\left\{\frac{\left(\frac{2\mu B}{\hbar^2}-\frac{\mu A}{\hbar^2\alpha}-\frac{\mu V_0}{2\hbar^2\alpha^2}+l(l+1)\right)+(n^2+n+\frac{1}{2})+(n+\frac{1}{2})\sqrt{\frac{8\mu B}{\hbar^2}+4l(l+1)+1}}{\left[1+2n+\sqrt{\frac{8\mu B}{\hbar^2}+4l(l+1)+1}\right]}\right\}+\frac{2\hbar^2\alpha^2}{\mu^2}\left\{\frac{\left(\frac{2\mu B}{\hbar^2}-\frac{\mu A}{\hbar^2\alpha}-\frac{\mu V_0}{2\hbar^2\alpha^2}+l(l+1)\right)+(n^2+n+\frac{1}{2})+(n+\frac{1}{2})\sqrt{\frac{8\mu B}{\hbar^2}+4l(l+1)+1}}{\left[1+2n+\sqrt{\frac{8\mu B}{\hbar^2}+4l(l+1)+1}\right]}\right\}^2 \qquad (59)
$$

However, taking the partial derivative of equation (51) with respect to $\mu$ gives

$$\left\langle \Psi(\mu) \right| \frac{\partial H(A)}{\partial A} \left| \Psi(\mu) \right\rangle = \frac{\hbar^2}{2\mu^2}\frac{d^2}{dr^2} - \frac{\hbar^2}{2\mu^2}\frac{l(l+1)}{r^2}$$

$$= \frac{1}{\mu}\left( \frac{\hbar^2}{2\mu}\frac{d^2}{dr^2} - \frac{\hbar^2}{2\mu}\frac{l(l+1)}{r^2} \right) \Rightarrow -\frac{1}{\mu}\left( \frac{\hbar^2}{2\mu}\frac{d^2}{dr^2} + \frac{\hbar^2}{2\mu}\frac{l(l+1)}{r^2} \right) \quad (60)$$

$$\Rightarrow -\frac{1}{\mu}\left(H - V\right) = -\frac{1}{\mu}\langle T \rangle$$

Hence,

$$\left\langle \Psi(\mu) \right| \frac{\partial H(A)}{\partial A} \left| \Psi(\mu) \right\rangle = \Rightarrow -\frac{1}{\mu}\left(H - V\right) = -\frac{1}{\mu}\langle T \rangle . \quad (61)$$

From the relation $T = \frac{p^2}{2\mu}$ , substituting for $T$ in equation (61) gives

$$\left\langle \Psi(\mu) \right| \frac{\partial H(A)}{\partial A} \left| \Psi(\mu) \right\rangle = \Rightarrow -\frac{1}{\mu}\left(H - V\right) = -\frac{1}{2\mu^2}\langle p^2 \rangle . \quad (62)$$

Equating equation (59) to (61) gives the expectation value of $\langle T \rangle$ . Therefore,

$$\langle T\rangle = 4\hbar^2\alpha^2 \left\{ \frac{\left[1+2n+\sqrt{\frac{8\mu B}{\hbar^2}+4l(l+1)+1}\right]\left[\frac{2B}{\hbar^2}-\frac{A}{\hbar^2\alpha}-\frac{v_0}{2\hbar^2\alpha^2}+\frac{\left(n+\frac{1}{2}\right)4B}{\hbar^2\sqrt{\frac{8\mu B}{\hbar^2}+4l(l+1)+1}}\right] - \left[\frac{\left(\frac{2\mu B}{\hbar^2}-\frac{\mu A}{\hbar^2\alpha}-\frac{\mu V_0}{2\hbar^2\alpha^2}+l(l+1)\right)4B}{\hbar^2\sqrt{\frac{8\mu B}{\hbar^2}+4l(l+1)+1}} + \frac{\left(n^2+n+\frac{1}{2}\right)(4B)}{\hbar^2\sqrt{\frac{8\mu B}{\hbar^2}+4l(l+1)+1}} + \frac{\left(n+\frac{1}{2}\right)(4B)}{\hbar^2}\right]}{\left[1+2n+\sqrt{\frac{8\mu B}{\hbar^2}+4l(l+1)+1}\right]^2} \right\} \times$$

$$\left\{ \frac{\left(\frac{2\mu B}{\hbar^2}-\frac{\mu A}{\hbar^2\alpha}-\frac{\mu V_0}{2\hbar^2\alpha^2}+l(l+1)\right) +(n^2+n+\frac{1}{2})+(n+\frac{1}{2})\sqrt{\frac{8\mu B}{\hbar^2}+4l(l+1)+1}}{\left[1+2n+\sqrt{\frac{8\mu B}{\hbar^2}+4l(l+1)+1}\right]} \right\} - \frac{2\hbar^2\alpha^2}{\mu}\left\{ \frac{\left(\frac{2\mu B}{\hbar^2}-\frac{\mu A}{\hbar^2\alpha}-\frac{\mu V_0}{2\hbar^2\alpha^2}+l(l+1)\right) +(n^2+n+\frac{1}{2})+(n+\frac{1}{2})\sqrt{\frac{8\mu B}{\hbar^2}+4l(l+1)+1}}{\left[1+2n+\sqrt{\frac{8\mu B}{\hbar^2}+4l(l+1)+1}\right]} \right\}^2 \quad (63)$$

Also, equating equating equation (59) to (62) gives $\langle p^2 \rangle$ .Thus,

$$\langle p^2\rangle = 8\hbar^2\alpha^2\mu\left\{\frac{\left[1+2n+\sqrt{\frac{8\mu B}{\hbar^2}+4l(l+1)+1}\right]\left[\frac{2B}{\hbar^2}-\frac{A}{\hbar^2\alpha}-\frac{v_0}{2\hbar^2\alpha^2}+\frac{\left(n+\frac{1}{2}\right)4B}{\hbar^2\sqrt{\frac{8\mu B}{\hbar^2}+4l(l+1)+1}}\right]-\left[\frac{\left(\frac{2\mu B}{\hbar^2}-\frac{\mu A}{\hbar^2\alpha}-\frac{\mu V_0}{2\hbar^2\alpha^2}+l(l+1)\right)4B}{\hbar^2\sqrt{\frac{8\mu B}{\hbar^2}+4l(l+1)+1}}+\frac{\left(n^2+n+\frac{1}{2}\right)(4B)}{\hbar^2\sqrt{\frac{8\mu B}{\hbar^2}+4l(l+1)+1}}+\frac{\left(n+\frac{1}{2}\right)(4B)}{\hbar^2}\right]}{\left[1+2n+\sqrt{\frac{8\mu B}{\hbar^2}+4l(l+1)+1}\right]^2}\right\}\times$$

$$\left\{\frac{\left(\frac{2\mu B}{\hbar^2}-\frac{\mu A}{\hbar^2\alpha}-\frac{\mu V_0}{2\hbar^2\alpha^2}+l(l+1)\right)+(n^2+n+\frac{1}{2})+(n+\frac{1}{2})\sqrt{\frac{8\mu B}{\hbar^2}+4l(l+1)+1}}{\left[1+2n+\sqrt{\frac{8\mu B}{\hbar^2}+4l(l+1)+1}\right]}\right\}-4\hbar^2\alpha^2\left\{\frac{\left(\frac{2\mu B}{\hbar^2}-\frac{\mu A}{\hbar^2\alpha}-\frac{\mu V_0}{2\hbar^2\alpha^2}+l(l+1)\right)+(n^2+n+\frac{1}{2})+(n+\frac{1}{2})\sqrt{\frac{8\mu B}{\hbar^2}+4l(l+1)+1}}{\left[1+2n+\sqrt{\frac{8\mu B}{\hbar^2}+4l(l+1)+1}\right]}\right\}^2 \qquad (64)$$

**SECTION 6: NUMERICAL SOLUTIONS FOR THE DIFFERENT EXPECTATION VALUES.**

Matlab algorithm is developed to solve numerically equations (55), (58), (63) and (64) for expectation values of $\langle r^{-2}\rangle, \langle r^{-1}\rangle, \langle T\rangle$ and $\langle p^2\rangle$ respectively for the different diatomic molecules. The diatomic molecules consider in this work are Hydrogen, Lithium hydride, hydrogen Chloride and carbon (ii) oxide. Table 1 is the spectroscopic constant used in the numerical computation of the expectation values .Tables 2-17 shows the numerical expectation values of $\langle r^{-2}\rangle, \langle r^{-1}\rangle, \langle T\rangle$ and $\langle p^2\rangle$ respectively for the four different diatomic molecules.. These tables show that some of the expectaion values increases with an increase in quantum state while some decreases with an increase in quantum state. Tables 2-5 shows the expectation values of $\langle r^{-2}\rangle$ for the diatomic molecules which decreases with an increase in quantum state for LiH, but increases with an increase in quantum state for $H_2$, HCl and CO respectively. The expectation value $\langle r^{-1}\rangle$ all decreases with an increase in quantum state for all the diatomic molecules as shown in tables 6-9. The expectation value $\langle T\rangle$ decreases with an increase in quantum state for $H_2$ and LiH respectively ; but increases with an increase in quantum state for HCl and CO as shown in tables 10-13. Finally, the expectation value $\langle p^2\rangle$ increases with an increase in quantum state for $H_2$ and CO but decreases with an increase in quantum state for LiH and HCl as shown in tables 14-17 respectively.

## Table 1 : Molecular Parameter Using Nikiforov-Uvarov Method

| Molecule | A($\dot{A}$) | B($\frac{1}{\dot{A}}$) | $C(\dot{A})$ | $\alpha$ | $\mu(amu)$ |
|---|---|---|---|---|---|
| Hydrogen($H_2$) | 0.7416 | 1.9426 | 1.440558 | 0.20990 | 0.5039100 |
| Lithium hydride($LiH$) | 1.5956 | 1.1280 | 1.7998368 | 1.55000 | 0.8801221 |
| Hydrogen Chloride($HCl$) | 1.2746 | 1.8677 | 2.38057 | 0.20039 | 0.9801045 |
| Carbon(II)oxide($CO$) | 1.1283 | 2.2994 | 2.59441 | 0.39000 | 6.8606719 |

**Table 2: Expectation values of $< r^{-2} >$ for hydrogen molecule**

| $n$ | $l$ | $< r^{-2} >_{nl}\ (\dot{A}^{-2})$ | $n$ | $l$ | $< r^{-2} >_{nl}\ (\dot{A}^{-2})$ | $n$ | $l$ | $< r^{-2} >_{nl}\ (\dot{A}^{-2})$ | $n$ | $l$ | $< r^{-2} >_{nl}\ (\dot{A}^{-2})$ |
|---|---|---|---|---|---|---|---|---|---|---|---|
| **0** | **0** | -2.03579269252 | **0** | **1** | -0.550065732697 | **0** | **2** | -0.100880019910 | **0** | **3** | -0.0119105671102 |
| **1** | **0** | -0.36962978563 | **1** | **1** | -0.101108401667 | **1** | **2** | -0.0120707336499 | **1** | **3** | -0.00003209757061 |
| **2** | **0** | -0.05920818720 | **2** | **1** | -0.009383143105 | **2** | **2** | -0.0002713032385 | **2** | **3** | -0.00702844221596 |
| **3** | **0** | -0.00123608985 | **3** | **1** | -0.001704790358 | **3** | **2** | -0.0106774421357 | **3** | **3** | -0.0201151142684 |
| **4** | **0** | -0.01046092114 | **4** | **1** | -0.019038983715 | **4** | **2** | -0.0282925702486 | **4** | **3** | -0.0351718504670 |
| **5** | **0** | -0.04084311251 | **5** | **1** | -0.04456442910 | **5** | **2** | -0.0482198274886 | **5** | **3** | -0.0507225834363 |
| **6** | **0** | -0.078170199119 | **6** | **1** | -0.07265739842 | **6** | **2** | -0.0686930232106 | **6** | **3** | -0.0662155593237 |
| **7** | **0** | -0.11746817864 | **7** | **1** | -0.10126588947 | **7** | **2** | -0.0890515939944 | **7** | **3** | -0.0814518703066 |
| **8** | **0** | -0.15688793153 | **8** | **1** | -0.12962320402 | **8** | **2** | -0.10905973352 | **8** | **3** | -0.0963742259414 |

**Table 3: Expectation values of $< r^{-2} >$ for Lithium hydride molecule**

| $n$ | $l$ | $< r^{-2} >_{nl}\ (\dot{A}^{-2})$ | $n$ | $l$ | $< r^{-2} >_{nl}\ (\dot{A}^{-2})$ | $n$ | $l$ | $< r^{-2} >_{nl}\ (\dot{A}^{-2})$ | $n$ | $l$ | $< r^{-2} >_{nl}\ (\dot{A}^{-2})$ |
|---|---|---|---|---|---|---|---|---|---|---|---|
| **0** | **0** | -2.50677160776 | **0** | **1** | -2.58048751355 | **0** | **2** | -2.59681215494 | **0** | **3** | -2.58530040749 |
| **1** | **0** | -4.33111511865 | **1** | **1** | -3.85673741137 | **1** | **2** | -3.48386144796 | **1** | **3** | -3.24729111386 |
| **2** | **0** | -6.05433406930 | **2** | **1** | -5.08754822400 | **2** | **2** | -4.35293111280 | **2** | **3** | -3.90141556913 |
| **3** | **0** | -7.73274556269 | **3** | **1** | -6.29620327700 | **3** | **2** | -5.21220036060 | **3** | **3** | -4.55084031801 |
| **4** | **0** | -9.38786532994 | **4** | **1** | -7.49251839380 | **4** | **2** | -6.06556567680 | **4** | **3** | -5.19723967523 |
| **5** | **0** | -11.0294166762 | **5** | **1** | -8.68128447320 | **5** | **2** | -6.91510709516 | **5** | **3** | -5.84157928177 |
| **6** | **0** | -12.6623937821 | **6** | **1** | -9.86510495707 | **6** | **2** | -7.76203290808 | **6** | **3** | -6.48445425671 |
| **7** | **0** | -14.2896150748 | **7** | **1** | -11.0455127066 | **7** | **2** | -8.60709195450 | **7** | **3** | -7.12625101366 |
| **8** | **0** | -15.9127886181 | **8** | **1** | -12.2234677505 | **8** | **2** | -9.45077263002 | **8** | **3** | -7.76723122881 |

**Table 4: Expectation values of $< r^{-2} >$ for Hydrogen Chloride molecule**

| $n$ | $l$ | $< r^{-2} >_{nl}\ (\dot{A}^{-2})$ | $n$ | $l$ | $< r^{-2} >_{nl}\ (\dot{A}^{-2})$ | $n$ | $l$ | $< r^{-2} >_{nl}\ (\dot{A}^{-2})$ | $n$ | $l$ | $< r^{-2} >_{nl}\ (\dot{A}^{-2})$ |
|---|---|---|---|---|---|---|---|---|---|---|---|
| **0** | **0** | -3.36841307312 | **0** | **1** | -1.49689154587 | **0** | **2** | -0.48709365309 | **0** | **3** | -0.146008021014 |
| **1** | **0** | -0.94639787381 | **1** | **1** | -0.455067253017 | **1** | **2** | -0.15349564684 | **1** | **3** | -0.041921828518 |
| **2** | **0** | -0.29836654065 | **2** | **1** | -0.141498891043 | **2** | **2** | -0.042378157709 | **2** | **3** | -0.0074968794792 |
| **3** | **0** | -0.08587628816 | **3** | **1** | -0.0355399041690 | **3** | **2** | -0.006430436462 | **3** | **3** | -0.0000019647915 |
| **4** | **0** | -0.01551154160 | **4** | **1** | -0.003439154420 | **4** | **2** | -0.0001287810650 | **4** | **3** | -0.004052078193 |
| **5** | **0** | -0.000010022621 | **5** | **1** | -0.001414035179 | **5** | **2** | -0.0066922782571 | **5** | **3** | -0.0133574528705 |
| **6** | **0** | -0.008007503911 | **6** | **1** | -0.0126923268912 | **6** | **2** | -0.0191475545351 | **6** | **3** | -0.0251076232366 |
| **7** | **0** | -0.026696710626 | **7** | **1** | -0.0301145909936 | **7** | **2** | -0.0343426512128 | **7** | **3** | -0.0379661955284 |
| **8** | **0** | -0.050359725356 | **8** | **1** | -0.0503830973850 | **8** | **2** | -0.0507669460586 | **8** | **3** | -0.0512710657544 |

**Table 5: Expectation values of $<r^{-2}>$ for Carbon(11)oxide molecule**

| $n$ | $l$ | $<r^{-2}>_{nl}\ (\dot{A}^{-2})$ | $n$ | $l$ | $<r^{-2}>_{nl}\ (\dot{A}^{-2})$ | $n$ | $l$ | $<r^{-2}>_{nl}\ (\dot{A}^{-2})$ | $n$ | $l$ | $<r^{-2}>_{nl}\ (\dot{A}^{-2})$ |
|---|---|---|---|---|---|---|---|---|---|---|---|
| **0** | **0** | -5.41245277040 | **0** | **1** | -4.72930366887 | **0** | **2** | -3.67655971494 | **0** | **3** | -2.61300040217 |
| **1** | **0** | -3.03066066748 | **1** | **1** | -2.66623256237 | **1** | **2** | -2.09567622314 | **1** | **3** | -1.50636102201 |
| **2** | **0** | -1.74783047367 | **2** | **1** | -1.54220867264 | **2** | **2** | -1.21700165807 | **2** | **3** | -0.876421321720 |
| **3** | **0** | -1.01689483720 | **3** | **1** | -0.896839504663 | **3** | **2** | -0.70588090076 | **3** | **3** | -0.504482166961 |
| **4** | **0** | -0.58507531875 | **4** | **1** | -0.513853152957 | **4** | **2** | -0.40039016144 | **4** | **3** | -0.280679157171 |
| **5** | **0** | -0.32518713046 | **5** | **1** | -0.283029567043 | **5** | **2** | -0.21607722006 | **5** | **3** | -0.145986805383 |
| **6** | **0** | -0.16881345832 | **6** | **1** | -0.144468886734 | **6** | **2** | -0.01061871058 | **6** | **3** | -0.0669337108866 |
| **7** | **0** | -0.077117801055 | **7** | **1** | -0.0638852202903 | **7** | **2** | -0.043549113783 | **7** | **3** | -0.236808524904 |
| **8** | **0** | -0.027061757436 | **8** | **1** | -0.0207992373230 | **8** | **2** | -0.011728402748 | **8** | **3** | 0.0040288991466 |

**Table 6: Expectation values of $<r^{-1}>$ for hydrogen molecule**

| $n$ | $l$ | $< r^{-1} >_{nl}\ (\dot{A}^{-1})$ | $n$ | $l$ | $< r^{-1} >_{nl}\ (\dot{A}^{-1})$ | $n$ | $l$ | $< r^{-1} >_{nl}\ (\dot{A}^{-1})$ | $n$ | $l$ | $< r^{-1} >_{nl}\ (\dot{A}^{-1})$ |
|---|---|---|---|---|---|---|---|---|---|---|---|
| **0** | **0** | 1.26036805460 | **0** | **1** | 0.679136007055 | **0** | **2** | 0.299286651245 | **0** | **3** | 0.104721107831 |
| **1** | **0** | 0.43797954010 | **1** | **1** | 0.246791348491 | **1** | **2** | 0.908972808251 | **1** | **3** | -0.0489334886369 |
| **2** | **0** | 0.151717261163 | **2** | **1** | 0.066410430386 | **2** | **2** | -0.122918741706 | **2** | **3** | -0.0663845048836 |
| **3** | **0** | 0.196001842741 | **3** | **1** | -0.0256311394072 | **3** | **2** | -0.707961120170 | **3** | **3** | -0.104294587266 |
| **4** | **0** | -0.052038753746 | **4** | **1** | -0.0788474858584 | **4** | **2** | -0.10713329778 | **4** | **3** | -0.129303455124 |
| **5** | **0** | -0.095181971633 | **5** | **1** | -0.112360323380 | **5** | **2** | -0.13123637757 | **5** | **3** | -0.146664218201 |
| **6** | **0** | -0.01231586721 | **6** | **1** | -0.134819740240 | **6** | **2** | -0.14803915423 | **6** | **3** | -0.159204967515 |
| **7** | **0** | -0.01423264246 | **7** | **1** | -0.150601728774 | **7** | **2** | -0.16021759087 | **7** | **3** | -0.168557899818 |
| **8** | **0** | -0.15602970503 | **8** | **1** | -0.162113050693 | **8** | **2** | -0.16932510673 | **8** | **3** | -0.175718563752 |

**Table 7: Expectation values of $< r^{-1} >$ for Lithium hydride molecule**

| $n$ | $l$ | $< r^{-1} >_{nl}\ (\dot{A}^{-1})$ | $n$ | $l$ | $< r^{-1} >_{nl}\ (\dot{A}^{-1})$ | $n$ | $l$ | $< r^{-1} >_{nl}\ (\dot{A}^{-1})$ | $n$ | $l$ | $< r^{-1} >_{nl}\ (\dot{A}^{-1})$ |
|---|---|---|---|---|---|---|---|---|---|---|---|
| **0** | **0** | -1.60040032851 | **0** | **1** | -1.68244122022 | **0** | **2** | -1.73643799656 | **0** | **3** | -1.76420127586 |
| **1** | **0** | -1.71692211724 | **1** | **1** | -1.74400394280 | **1** | **2** | -1.76620141352 | **1** | **3** | -1.77988044276 |
| **2** | **0** | -1.75762881150 | **2** | **1** | -1.76973455508 | **2** | **2** | -1.78095151220 | **2** | **3** | -1.78867975015 |
| **3** | **0** | -1.77645171927 | **3** | **1** | -1.78287756416 | **3** | **2** | -1.78931865180 | **3** | **3** | -1.79410620876 |
| **4** | **0** | -1.78667027824 | **4** | **1** | -1.79048175684 | **4** | **2** | -1.79451743236 | **4** | **3** | -1.79768675018 |
| **5** | **0** | -1.79282918066 | **5** | **1** | -1.79527280193 | **5** | **2** | -1.79796681743 | **5** | **3** | -1.80017271446 |
| **6** | **0** | -1.79682533350 | **6** | **1** | -1.79848479954 | **6** | **2** | -1.80037196611 | **6** | **3** | -1.80196871384 |
| **7** | **0** | -1.79956445077 | **7** | **1** | -1.80074247145 | **7** | **2** | -1.80211547968 | **7** | **3** | -1.80330831803 |
| **8** | **0** | -1.80152336966 | **8** | **1** | -1.80238957942 | **8** | **2** | -1.80341952627 | **8** | **3** | -1.80433401747 |

**Table 8: Expectation values of $< r^{-1} >$ for hydrogen chloride molecule**

| $n$ | $l$ | $< r^{-1} >_{nl}$ $(\dot{A}^{-1})$ | $n$ | $l$ | $< r^{-1} >_{nl}$ $(\dot{A}^{-1})$ | $n$ | $l$ | $< r^{-1} >_{nl}$ $(\dot{A}^{-1})$ | $n$ | $l$ | $< r^{-1} >_{nl}$ $(\dot{A}^{-1})$ |
|---|---|---|---|---|---|---|---|---|---|---|---|
| **0** | **0** | 1.67290061760 | **0** | **1** | 1.13680691238 | **0** | **2** | 0.661455723646 | **0** | **3** | 0.367491764061 |
| **1** | **0** | 0.74846462792 | **1** | **1** | 0.541241188088 | **1** | **2** | 0.328957147400 | **1** | **3** | 0.178076414067 |
| **2** | **0** | 0.37036272045 | **2** | **1** | 0.269473851341 | **2** | **2** | 0.156801277037 | **2** | **3** | 0.0692601602849 |
| **3** | **0** | 0 .17964566098 | **3** | **1** | 0.123141317091 | **3** | **2** | 0.056299132841 | **3** | **3** | 0.00104370523532 |
| **4** | **0** | 0.070209512012 | **4** | **1** | 0.035435531765 | **4** | **2** | -0.007428136083 | **4** | **3** | -0.0445186237639 |
| **5** | **0** | 0.016610598991 | **5** | **1** | -0.021241162011 | **5** | **2** | -0.050356733813 | **5** | **3** | -0.0764508536791 |
| **6** | **0** | -0.440948649449 | **6** | **1** | -0.059969931387 | **6** | **2** | -0.080643040971 | **6** | **3** | -0.0996933551051 |
| **7** | **0** | -0.076147510665 | **7** | **1** | -0.087600307593 | **7** | **2** | -0.10280451060 | **7** | **3** | -0.117135259492 |
| **8** | **0** | -0.099468802646 | **8** | **1** | -0.108000723652 | **8** | **2** | -0.11950730360 | **8** | **3** | -0.130557661145 |

**Table 9: Expectation values of $< r^{-1} >$ for carbon (11) oxide molecule**

| $n$ | $l$ | $< r^{-1} >_{nl}$ $(\dot{A}^{-1})$ | $n$ | $l$ | $< r^{-1} >_{nl}$ $(\dot{A}^{-1})$ | $n$ | $l$ | $< r^{-1} >_{nl}$ $(\dot{A}^{-1})$ | $n$ | $l$ | $< r^{-1} >_{nl}$ $(\dot{A}^{-1})$ |
|---|---|---|---|---|---|---|---|---|---|---|---|
| **0** | **0** | 2.31839526263 | **0** | **1** | 2.16980581181 | **0** | **2** | 1.91725711912 | **0** | **3** | 1.62066078451 |
| **1** | **0** | 1.60875651016 | **1** | **1** | 1.51371944413 | **1** | **2** | 1.34956632094 | **1** | **3** | 1.15223807283 |
| **2** | **0** | 1.14420890115 | **2** | **1** | 1.07979143977 | **2** | **2** | 0.967151726606 | **2** | **3** | 0.829295929385 |
| **3** | **0** | 0.82362691792 | **3** | **1** | 0.777968743869 | **3** | **2** | 0.697356029706 | **3** | **3** | 0.597281349608 |
| **4** | **0** | 0.59313111040 | **4** | **1** | 0.559601038803 | **4** | **2** | 0.499937457666 | **4** | **3** | 0.425006485067 |
| **5** | **0** | 0.42187751753 | **5** | **1** | 0.396533698615 | **5** | **2** | 0.351145600465 | **5** | **3** | 0.293592864040 |
| **6** | **0** | 0.29117578027 | **6** | **1** | 0.271556148258 | **6** | **2** | 0.236229286498 | **6** | **3** | 0.191070266067 |
| **7** | **0** | 0.18916450089 | **7** | **1** | 0.173667189339 | **7** | **2** | 0.145634487129 | **7** | **3** | 0.109551116308 |
| **8** | **0** | 0.10802204395 | **8** | **1** | 0.095568524738 | **8** | **2** | 0.072952403977 | **8** | **3** | 0.0436667613675 |

**Table 10: Expectation values of $<T>$ for hydrogen molecule**

| $n$ | $l$ | $<T>_{nl}\ (eV)$ | $n$ | $l$ | $<T>_{nl}\ (eV)$ | $n$ | $l$ | $<T>_{nl}\ (eV)$ | $n$ | $l$ | $<T>_{nl}\ (eV)$ |
|---|---|---|---|---|---|---|---|---|---|---|---|
| **0** | **0** | -5.77750109574 | **0** | **1** | -1.13052963585 | **0** | **2** | 0.394536773578 | **0** | **3** | 0.423197715186 |
| **1** | **0** | -0.1575248128 | **1** | **1** | 0.406254338351 | **1** | **2** | 0.376555036342 | **1** | **3** | -0.033193807081 |
| **2** | **0** | 0.38336476605 | **2** | **1** | 0.290302301857 | **2** | **2** | -0.08429460330 | **2** | **3** | -.640317976640 |
| **3** | **0** | 0.10046884400 | **3** | **1** | -0.180643611080 | **3** | **2** | -0.68724169821 | **3** | **3** | -1.32938786534 |
| **4** | **0** | -0.40517604920 | **4** | **1** | -0.779382166318 | **4** | **2** | -1.37160329806 | **4** | **3** | -2.08792109767 |
| **5** | **0** | -1.01427575812 | **5** | **1** | -1.45846545276 | **5** | **2** | -2.12631456936 | **5** | **3** | -2.91618602337 |
| **6** | **0** | -1.70090079690 | **6** | **1** | -2.20927404251 | **6** | **2** | -2.95172760482 | **6** | **3** | -3.81704612613 |
| **7** | **0** | -2.46092418669 | **7** | **1** | -3.03235962906 | **7** | **2** | -3.85053948565 | **7** | **3** | -4.79344101943 |
| **8** | **0** | -3.29551550979 | **8** | **1** | -3.93017085262 | **8** | **2** | -4.82551066701 | **8** | **3** | -5.84782747460 |

**Table 11: Expectation values of $<T>$ for Lithium hydride molecule**

| $n$ | $l$ | $<T>_{nl}\ (eV)$ | $n$ | $l$ | $<T>_{nl}\ (eV)$ | $n$ | $l$ | $<T>_{nl}\ (eV)$ | $n$ | $l$ | $<T>_{nl}\ (eV)$ |
|---|---|---|---|---|---|---|---|---|---|---|---|
| **0** | **0** | -4.23055856747 | **0** | **1** | -7.80886692925 | **0** | **2** | -14.2391535700 | **0** | **3** | -23.4729539098 |
| **1** | **0** | -8.53985347030 | **1** | **1** | -14.2714091244 | **1** | **2** | -23.4467711704 | **1** | **3** | -35.4898681236 |
| **2** | **0** | -15.8689882670 | **2** | **1** | -23.6854511210 | **2** | **2** | -35.5261470800 | **2** | **3** | -50.3254081811 |
| **3** | **0** | -26.0744836125 | **3** | **1** | -35.9457058061 | **3** | **2** | -50.4157170230 | **3** | **3** | -67.9449624090 |
| **4** | **0** | -39.0894871134 | **4** | **1** | -51.0026938513 | **4** | **2** | -68.0844826720 | **4** | **3** | -88.3296507196 |
| **5** | **0** | -54.8817584651 | **5** | **1** | -68.8311306737 | **5** | **2** | -88.5154126054 | **5** | **3** | -111.468380478 |
| **6** | **0** | -73.4342035101 | **6** | **1** | -89.4169299119 | **6** | **2** | -111.698439709 | **6** | **3** | -137.354231838 |
| **7** | **0** | -94.7369941855 | **7** | **1** | -112.751668968 | **7** | **2** | -137.627252744 | **7** | **3** | -165.982673447 |
| **8** | **0** | -118.784101859 | **8** | **1** | -138.830017754 | **8** | **2** | -166.297702040 | **8** | **3** | -197.350617564 |

**Table 12: Expectation values of $<T>$ for hydrogen chloride molecule**

| $n$ | $l$ | $<T>_{nl}\ (eV)$ | $n$ | $l$ | $<T>_{nl}\ (eV)$ | $n$ | $l$ | $<T>_{nl}\ (eV)$ | $n$ | $l$ | $<T>_{nl}\ (eV)$ |
|---|---|---|---|---|---|---|---|---|---|---|---|
| **0** | **0** | -16.8602470534 | **0** | **1** | -1.82598811792 | **0** | **2** | -0.374986529886 | **0** | **3** | 0.187962968533 |
| **1** | **0** | -3.81737409389 | **1** | **1** | -0.165259693767 | **1** | **2** | 0.198272886276 | **1** | **3** | 0.225715690749 |
| **2** | **0** | -0.72649687996 | **2** | **1** | 0.205070417862 | **2** | **2** | 0.192615414252 | **2** | **3** | 0.005605007516 |
| **3** | **0** | 0.81708122726 | **3** | **1** | 0.133144274232 | **3** | **2** | -0.040900866930 | **3** | **3** | -0.333574338340 |
| **4** | **0** | 0.18087961451 | **4** | **1** | -0.123394560842 | **4** | **2** | -0.38340136182 | **4** | **3** | -0.740549288126 |
| **5** | **0** | 0.00764339440 | **5** | **1** | -0.473864241728 | **5** | **2** | -0.79022687896 | **5** | **3** | -1.19515943846 |
| **6** | **0** | -0.29259122490 | **6** | **1** | -0.883279654269 | **6** | **2** | -1.24360828824 | **6** | **3** | -1.68935744081 |
| **7** | **0** | -0.66544419046 | **7** | **1** | -1.33747314334 | **7** | **2** | -1.73639795700 | **7** | **3** | -2.22010238289 |
| **8** | **0** | -1.08910266100 | **8** | **1** | -2.75189176950 | **8** | **2** | -0.52753030521 | **8** | **3** | 0.187962968533 |

**Table 13: Expectation values of $<T>$ for carbon (11) oxide molecule**

| $n$ | $l$ | $<T>_{nl}\ (eV)$ | $n$ | $l$ | $<T>_{nl}\ (eV)$ | $n$ | $l$ | $<T>_{nl}\ (eV)$ | $n$ | $l$ | $<T>_{nl}\ (eV)$ |
|---|---|---|---|---|---|---|---|---|---|---|---|
| **0** | **0** | -129.929113083 | **0** | **1** | -117.634020526 | **0** | **2** | -97.7238969338 | **0** | **3** | -76.0158584874 |
| **1** | **0** | -75.6124342843 | **1** | **1** | -69.0411092188 | **1** | **2** | -58.1864276360 | **1** | **3** | -46.0107012922 |
| **2** | **0** | -45.8529997998 | **2** | **1** | -42.0768329533 | **2** | **2** | -35.7490746782 | **2** | **3** | -28.5036356479 |
| **3** | **0** | -28.4567951051 | **3** | **1** | -26.1683200430 | **3** | **2** | -22.2936665438 | **3** | **3** | -17.7913843239 |
| **4** | **0** | -17.7933612549 | **4** | **1** | -16.3516907261 | **4** | **2** | -13.8933817128 | **4** | **3** | -11.0086725080 |
| **5** | **0** | -11.0290843042 | **5** | **1** | -10.0962337215 | **5** | **2** | -8.49886386390 | **5** | **3** | -6.61454246456 |
| **6** | **0** | -6.63784565004 | **6** | **1** | -6.02477650635 | **6** | **2** | -4.97370747443 | **6** | **3** | -3.73341287233 |
| **7** | **0** | -3.75143695048 | **7** | **1** | -3.34718744046 | **7** | **2** | -2.65576948817 | **7** | **3** | -1.84472534278 |
| **8** | **0** | -1.85313324544 | **8** | **1** | -1.58989078767 | **8** | **2** | -1.14303953741 | **8** | **3** | -0.627040547989 |

**Table 14: Expectation values of $< P^2 >$ for hydrogen molecule**

| $n$ | $l$ | $< p^2 >_{nl} \left(\frac{eV}{c}\right)^2$ | $n$ | $l$ | $< p^2 >_{nl} \left(\frac{eV}{c}\right)^2$ | $n$ | $l$ | $< p^2 >_{nl} \left(\frac{eV}{c}\right)^2$ | $n$ | $l$ | $< p^2 >_{nl} \left(\frac{eV}{c}\right)^2$ |
|---|---|---|---|---|---|---|---|---|---|---|---|
| **0** | **0** | -5.8226811543 | **0** | **1** | -1.13937037760 | **0** | **2** | 0.394536773578 | **0** | **3** | .426507121319 |
| **1** | **0** | -0.1587566569 | **1** | **1** | 0.409431247278 | **1** | **2** | 0.379499696727 | **1** | **3** | -0.033453382653 |
| **2** | **0** | 0.38636267852 | **2** | **1** | 0.292572465858 | **2** | **2** | -0.849537871017 | **2** | **3** | -0.645325263217 |
| **3** | **0** | 0.10125451036 | **3** | **1** | -0.180643611080 | **3** | **2** | -0.69261592829 | **3** | **3** | -1.33978367845 |
| **4** | **0** | -0.40834452590 | **4** | **1** | -0.785476934860 | **4** | **2** | -1.38232923585 | **4** | **3** | -2.10424864066 |
| **5** | **0** | -1.02220739455 | **5** | **1** | -1.46987065261 | **5** | **2** | -2.14294234930 | **5** | **3** | -2.93899059808 |
| **6** | **0** | -1.71420184114 | **6** | **1** | -2.22655056553 | **6** | **2** | -2.97481011469 | **6** | **3** | -3.84689542683 |
| **7** | **0** | -2.48016861384 | **7** | **1** | -3.05607268136 | **7** | **2** | -3.88065070443 | **7** | **3** | -4.83092572821 |
| **8** | **0** | -3.32128644108 | **8** | **1** | -3.96090478869 | **8** | **2** | -4.86324616043 | **8** | **3** | -5.89355748546 |

**Table 15: Expectation values of $< p^2 >$ for Lithium hydride molecule**

| $n$ | $l$ | $< p^2 >_{nl} \left(\frac{eV}{c}\right)^2$ | $n$ | $l$ | $< p^2 >_{nl} \left(\frac{eV}{c}\right)^2$ | $n$ | $l$ | $< p^2 >_{nl} \left(\frac{eV}{c}\right)^2$ | $n$ | $l$ | $< p^2 >_{nl} \left(\frac{eV}{c}\right)^2$ |
|---|---|---|---|---|---|---|---|---|---|---|---|
| **0** | **0** | -7.44681618115 | **0** | **1** | 13.7455127207 | **0** | **2** | -14.2391535700 | **0** | **3** | -41.3181309765 |
| **1** | **0** | -15.0322275400 | **1** | **1** | -25.1211651371 | **1** | **2** | -41.2720429614 | **1** | **3** | -62.4708345234 |
| **2** | **0** | -27.9332945570 | **2** | **1** | -41.6921779602 | **2** | **2** | -62.5346943459 | **2** | **3** | -88.5850078633 |
| **3** | **0** | -45.8974585468 | **3** | **1** | -35.9457058061 | **3** | **2** | -88.7439734787 | **3** | **3** | -119.599726000 |
| **4** | **0** | -68.8070429723 | **4** | **1** | -89.7771960362 | **4** | **2** | -119.845315733 | **4** | **3** | -155.481755367 |
| **5** | **0** | -96.6052970237 | **5** | **1** | -121.159598548 | **5** | **2** | -155.808741649 | **5** | **3** | -196.211570220 |
| **6** | **0** | -129.262130810 | **6** | **1** | -157.395632259 | **6** | **2** | -196.616530647 | **6** | **3** | -241.776989938 |
| **7** | **0** | -166.760244540 | **7** | **1** | -198.470471341 | **7** | **2** | -242.257573406 | **7** | **3** | -292.170038235 |
| **8** | **0** | -209.089026350 | **8** | **1** | -244.374733538 | **8** | **2** | -292.724565489 | **8** | **3** | -347.385279935 |

**Table 16: Expectation values of $< p^2 >$ for hydrogen chloride molecule**

| $n$ | $l$ | $< p^2 >_{nl} \left(\frac{eV}{c}\right)^2$ | $n$ | $l$ | $< p^2 >_{nl} \left(\frac{eV}{c}\right)^2$ | $n$ | $l$ | $< p^2 >_{nl} \left(\frac{eV}{c}\right)^2$ | $n$ | $l$ | $< p^2 >_{nl} \left(\frac{eV}{c}\right)^2$ |
|---|---|---|---|---|---|---|---|---|---|---|---|
| **0** | **0** | -7.44681618115 | **0** | **1** | -13.7455127207 | **0** | **2** | -14.2391535700 | **0** | **3** | -41.3181309765 |
| **1** | **0** | -15.0322275400 | **1** | **1** | -25.1211651371 | **1** | **2** | -41.2720429614 | **1** | **3** | -62.4708345234 |
| **2** | **0** | -27.9332945570 | **2** | **1** | -41.6921779602 | **2** | **2** | -62.5346943459 | **2** | **3** | -88.5850078633 |
| **3** | **0** | -45.8974585468 | **3** | **1** | -35.9457058061 | **3** | **2** | -88.7439734787 | **3** | **3** | -119.599726000 |
| **4** | **0** | -68.8070429723 | **4** | **1** | -89.7771960362 | **4** | **2** | -119.845315733 | **4** | **3** | -155.481755367 |
| **5** | **0** | -96.6052970237 | **5** | **1** | -121.159598548 | **5** | **2** | -155.808741649 | **5** | **3** | -196.211570220 |
| **6** | **0** | -129.262130810 | **6** | **1** | -157.395632259 | **6** | **2** | -196.616530647 | **6** | **3** | -241.776989938 |
| **7** | **0** | -166.760244540 | **7** | **1** | -198.470471341 | **7** | **2** | -242.257573406 | **7** | **3** | -292.170038235 |
| **8** | **0** | -209.089026350 | **8** | **1** | -244.374733538 | **8** | **2** | -292.724565489 | **8** | **3** | -347.385279935 |

**Table 17: Expectation values of $< p^2 >$ for carbon (11) oxide molecule**

| $n$ | $l$ | $< p^2 >_{nl} \left(\frac{eV}{c}\right)^2$ | $n$ | $l$ | $< p^2 >_{nl} \left(\frac{eV}{c}\right)^2$ | $n$ | $l$ | $< p^2 >_{nl} \left(\frac{eV}{c}\right)^2$ | $n$ | $l$ | $< p^2 >_{nl} \left(\frac{eV}{c}\right)^2$ |
|---|---|---|---|---|---|---|---|---|---|---|---|
| **0** | **0** | -1782.80203024 | **0** | **1** | -1614.09683822 | **0** | **2** | -97.7238969338 | **0** | **3** | -1043.03972856 |
| **1** | **0** | -1037.50420637 | **1** | **1** | -947.336795925 | **1** | **2** | -798.395978086 | **1** | **3** | -631.328650911 |
| **2** | **0** | -629.164774514 | **2** | **1** | -577.350690967 | **2** | **2** | -490.525344190 | **2** | **3** | -391.108184274 |
| **3** | **0** | -390.465469082 | **3** | **1** | -26.1683200430 | **3** | **2** | -305.899063210 | **3** | **3** | -244.121700986 |
| **4** | **0** | -244.148827137 | **4** | **1** | -224.367170164 | **4** | **2** | -190.635867026 | **4** | **3** | -151.053780264 |
| **5** | **0** | -151.333857537 | **5** | **1** | -138.533893977 | **5** | **2** | -116.615832986 | **5** | **3** | -90.7604112360 |
| **6** | **0** | -91.0801622557 | **6** | **1** | -82.6680297618 | **6** | **2** | -68.2459502172 | **6** | **3** | -51.2274415685 |
| **7** | **0** | -51.4747561415 | **7** | **1** | -45.9279096337 | **7** | **2** | -36.4407262008 | **7** | **3** | -25.3121106450 |
| **8** | **0** | -25.4274783679 | **8** | **1** | -21.8154381022 | **8** | **2** | -15.6840384698 | **8** | **3** | -8.60383893549 |

**SECTION 7: CONCLUSION**

In this work, we obtained analytical solution of Schrodinger equation with quantum interaction potential (HYIQP)and obtained the energy eigen value and the wave function using conventional Nikiforov-Uvarov method. Under some special cases, the potential reduced to three well known potentials which are: Hulthen, Yukawa and inversely quadratic potential. We further compute the expectation values for four diatomic molecules using Hellmann Feynman theorem. We obtained the numerical solutions of the expectation values by implementing Matlab algorithm. We implement mathematica algorithm to obtain the wave function and probability density plots for four diatomic molecules with an increasing orbital angular momentum. The quantum interaction potential model developed is suitable in studying bound state energies of diatomic molecules. Figure 1 shows the graph of the quantum interaction potential with different values of the screening parameter. However, figure 2 shows the graph of the potential (F) plotted against the individual potentials (F1,F2 and F3) . Some of the numerical expectation values either increase or decrease with an increase in quantum state.

The authors declare that there is no conflict of interest regarding the publication of this paper.